\documentclass[sn-mathphys,Numbered]{sn-jnl}

\usepackage{graphicx}
\usepackage{multirow}
\usepackage{amsmath,amssymb,amsfonts}
\usepackage{amsthm}
\usepackage{mathrsfs}
\usepackage[title]{appendix}
\usepackage{xcolor}
\usepackage{textcomp}
\usepackage{manyfoot}
\usepackage{booktabs}
\usepackage{algorithm}
\usepackage{algorithmicx}
\usepackage{algpseudocode}
\usepackage{listings}

\numberwithin{equation}{section}

\newcommand{\beq}{\begin{equation}}
\newcommand{\eeq}{\end{equation}}
\newcommand{\be}{\begin{equation*}}
\newcommand{\ee}{\end{equation*}}
\newcommand{\beqa}{\begin{eqnarray}}
\newcommand{\eeqa}{\end{eqnarray}}
\newcommand{\bea}{\begin{eqnarray*}}
\newcommand{\eea}{\end{eqnarray*}}

\renewcommand{\aa}{{\alpha}}
\newcommand{\abs}[1]{\vert#1\vert}
\newcommand{\cc}{{\cal C}}
\newcommand{\dd}{{\rm d}}
\newcommand{\eps}{{\varepsilon}}
\newcommand{\eq}{\mathrm{eq}}

\newcommand{\erfc}{\mathop{\rm erfc}}
\newcommand{\even}{{\rm even}}
\newcommand{\frad}[2]{{\displaystyle{\displaystyle#1\over\displaystyle#2}}}
\newcommand{\ii}{{\rm i}}
\newcommand{\lam}{{\lambda}}
\newcommand{\lap}[1]{\mathrel{\mathop{\cal L}\limits_{#1}^{}}}

\renewcommand{\max}{{\rm max}}
\renewcommand{\min}{{\rm min}}
\newcommand{\mean}[1]{\langle#1\rangle}
\newcommand{\nt}{{\mathcal N}_t}
\newcommand{\odd}{{\rm odd}}
\renewcommand{\phi}{{\varphi}}
\newcommand{\prob}{\mathbb{P}}
\newcommand{\sg}{{\rm sg}}
\renewcommand{\th}{{\theta}}
\newcommand{\var}{\mathop{\rm Var}\nolimits}
\newcommand{\vsk}{\vskip4pt\noindent}
\newcommand{\z}{{\zeta}}
\renewcommand{\L}{{\cal L}}
\newcommand{\R}{{\cal R}}
\newcommand{\Z}{{\cal Z}}

\newcommand{\X}{X_\th}
\newcommand{\Xone}{X_{\th_1}}
\newcommand{\Xtwo}{X_{\th_2}}
\newcommand{\Y}{Y_{\th_1,\th_2}}
 \newcommand{\T}{{\boldsymbol{T}}} 
 \newcommand{\ttau}{{\boldsymbol{\tau}}} 
\newcommand{\taum}{\mean{\ttau}}
\newcommand{\ttaur}{{\boldsymbol\tau}^{(\tt r)}}

\newcommand{\e}{{\rm e}}

\begin{document}

\title{Replicating a renewal process at random times}

\author*[]{\fnm{Claude} \sur{Godr\`eche*}}\email{claude.godreche@ipht.fr}

\author[]{\fnm{Jean-Marc} \sur{Luck}}\email{jean-marc.luck@ipht.fr}

\affil[]{
\orgdiv{Universit\'e Paris-Saclay, CEA, CNRS}, 
\orgname{Institut de Physique Th\'eorique}, 
\postcode{91191}
\city{Gif-sur-Yvette}, 
\country{France}}

\abstract{
We replicate a renewal process at random times, which is equivalent to nesting two renewal processes, or considering a renewal process subject to stochastic resetting.
We investigate the consequences on the statistical properties of the model of the intricate interplay between the two probability laws
governing the distribution of time intervals between renewals, on the one hand,
and of time intervals between resettings, on the other hand.
In particular, the total number ${\mathcal N}_t$ of renewal events occurring within a specified observation time exhibits a remarkable range of behaviours,
depending on the exponents characterising the power-law decays of the two
probability distributions.
Specifically, ${\mathcal N}_t$ can either grow linearly in time
and have relatively negligible fluctuations,
or grow subextensively over time while continuing to fluctuate.
These behaviours highlight the dominance of the most regular process across all regions of the phase diagram.
In the presence of Poissonian resetting, the statistics of ${\mathcal N}_t$
is described by a unique `dressed' renewal process, which is a deformation of the
renewal process without resetting.
We also discuss the relevance of the present study to first passage under restart and to continuous time random walks subject to stochastic resetting.
}




\maketitle
\section{Introduction}
\label{sec:intro}

A renewal process is a stochastic model in which events occur randomly over
time, resetting the clock for the next event.
The interarrival times between events are independent and identically
distributed (iid) random variables with a common arbitrary distribution.
The Poisson process, which corresponds to choosing an exponential distribution
of interarrival times, is the simplest example of a renewal process~\cite{feller1,feller2,cox,cox-miller,grimmett}.

\begin{figure}[h]
\centering
\includegraphics[angle=0,width=0.9\linewidth,clip=true]{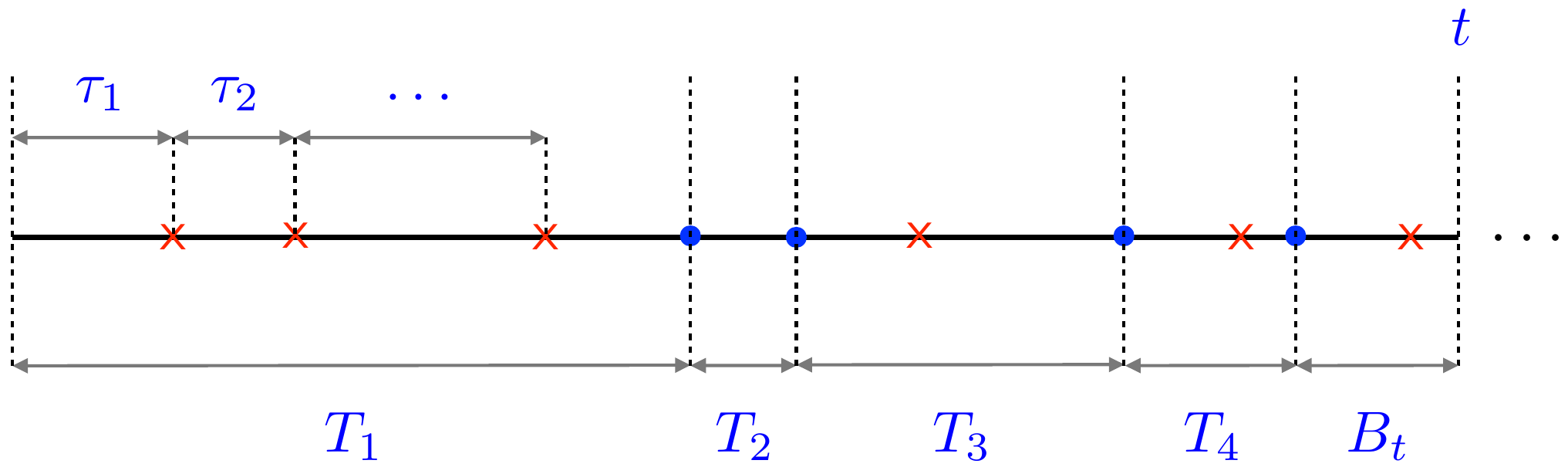}
\caption{\small
An example of two nested renewal processes with
five cycles of replication of the internal renewal process.
Renewal events of the internal process are figured by crosses, replication
events (or resettings) due to the external renewal process by dots.
The intervals of time between two crosses, $\ttau_1,\ttau_2,\dots$, have common
probability density $\rho(\tau)$.
The intervals of time between two dots, $\T_1,\T_2,\dots$, have common
probability density $f(T)$.
The last interval, $B_t$, represents the backward recurrence time or the age of
the external process at time $t$, which indicates the time elapsed since the
last replication (or resetting).
In this example, the total number $\nt$ of internal renewals (i.e., of crosses)
is equal to $6$.
}
\label{fig:nested}
\end{figure}

In this work, we investigate a theoretical model
consisting of two nested renewal processes.
The first one---dubbed the internal process---is replicated at random intervals of time,
drawn from a distribution characterising the second one---dubbed the external process.
The probability density of interarrival times of the internal process will be
denoted by $\rho(\tau)$, and that of the external process by $f(T)$.
An illustration is provided in figure~\ref{fig:nested},
which depicts five cycles of replication of the internal process,
of respective durations $\T_1,\dots,\T_4$ and $B_t$.
The last interval, $B_t$, represents the backward recurrence time,
or age of the external process at time $t$,
which is the time elapsed since the last replication event.

To provide a concrete example, in a manufacturing setting, the two nested
renewal processes would correspond respectively to the intervals of time
between component failures (internal process) and the intervals of time between component replacements (external process).
In the context of reliability analysis, the concept of nested renewal processes
has been previously introduced in~\cite{ansell}, with the following definitions.
\textit{Shocks occur to a component randomly in time in an ordinary renewal process, each shock causing a random amount of damage. 
Damages are identically and independently distributed, and damages resulting from shocks are accumulated. 
In addition to this cumulative process there is a second ordinary renewal process in time the effect of which is to restart the cumulative renewal process at zero accumulated shocks and consequently zero cumulative damage. 
This represents component replacement.}
This reference will be further examined later.
In a broader context, the idea of nesting stochastic processes (not limited to renewal
processes) across different scales has been investigated in other disciplines,
including the stochastic modeling of precipitation.
For example, in~\cite{paschalis}, an external model is employed to represent
the processes related to storm occurrences and the time periods between them,
while an internal model nested within it is utilised to capture the variability
of rainfall within a given storm.

Interestingly, the model described above, involving two nested renewal
processes, happens to be a specific instance of 
a class of models which has been recently popularised under the name of stochastic processes
under resetting.
Processes of this type have been studied for a long time,
as documented in~\cite{montero2017} for a historical perspective.
Lately, these models have gained significant attention in the field of statistical physics
(see~\cite{EMS,gupta} for reviews).
One notable aspect of the present study is that the stochastic process
subject to resetting is, in fact, a renewal process itself.
This (internal) renewal process, characterised by the density $\rho(\tau)$, is
reset at random time intervals, which are drawn from the density $f(T)$, characterising the external renewal process.

A simple example of such a process is naturally encountered when
considering the simple random walk with steps $\pm1$ (or P\'olya walk~\cite{polya}) on the
one-dimensional lattice, subject to stochastic resetting\footnote{For reference, other aspects of the P\'olya walk, or of more general lattice random walks, subject to stochastic resetting, have been explored in~\cite{majumdar2015,bonomo,glmaxpol,kumar}.}.
The events of the internal process are the epochs of the returns to the origin
of the walk, while the events of the external process are the reset events in
discrete time, corresponding to restarting the walk with a given probability.
A companion paper will be entirely dedicated to the study of this process~\cite{glpolya}.
Another example where such a process is encountered is when considering the continuous time random walk under resetting, as will be commented upon later (see sections~\ref{sec:over} and~\ref{sec:NBt}).

In this paper, we consider the more abstract model of nested renewal processes in full generality, which implies that we shall allow
the two distributions associated with the internal and external renewal
processes to be arbitrary\footnote{Reset stochastic processes with arbitrary
distributions of the time intervals between resettings have been considered in prior works such as~\cite{nagar,eule,reuveni,bodrova,shkilev,mishra,barkai2023}.}.
We shall be mostly interested in the case where these distributions have
power-law tails with respective exponents $\th_1$ and $\th_2$ (see~(\ref{eq:t1t2})).

\begin{figure}
\begin{center}
\includegraphics[angle=0,width=0.6\linewidth,clip=true]{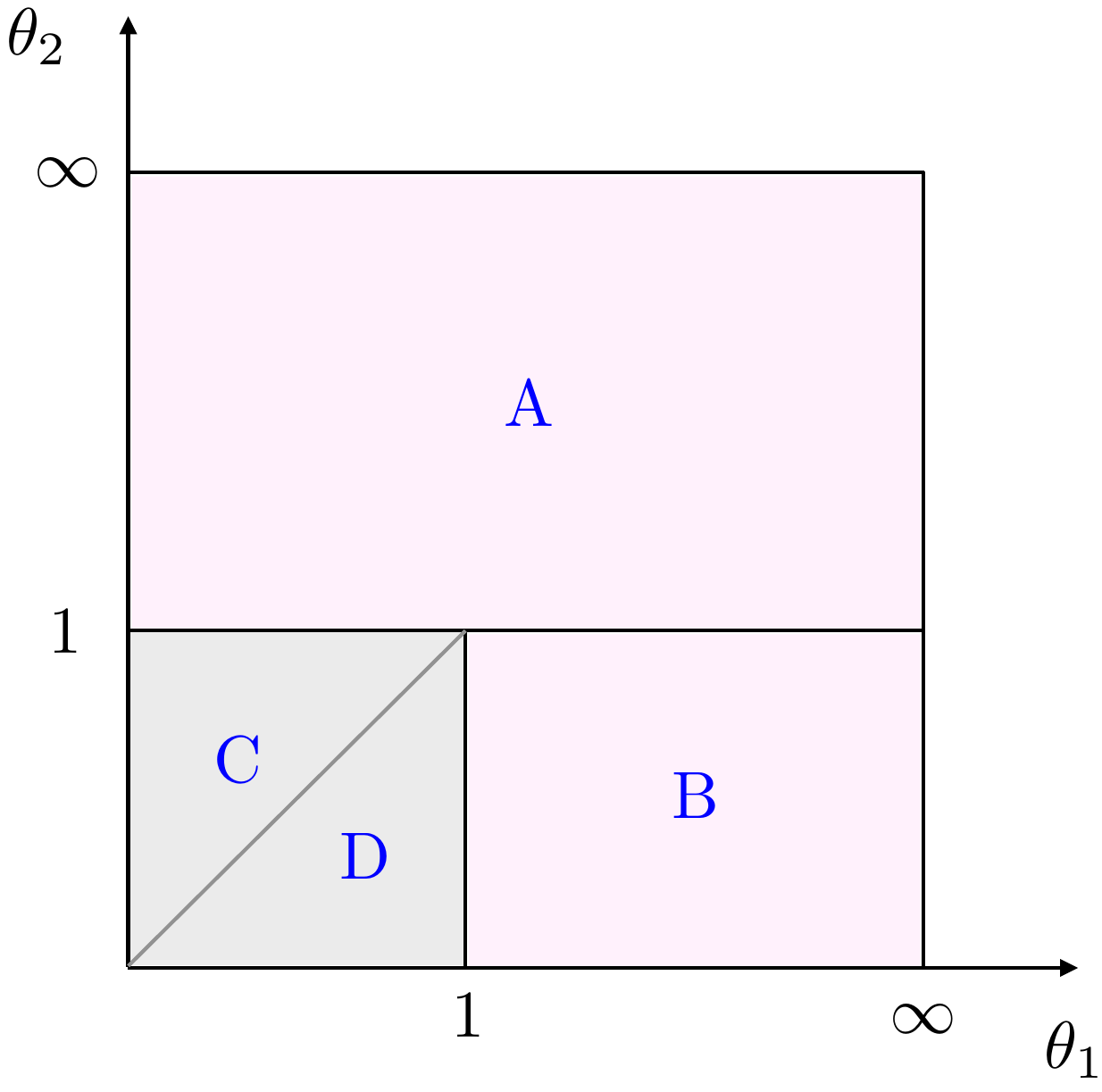}
\caption{\small
The four different regions of the phase diagram in the $\th_1$--$\th_2$-plane.
In regions~A and~B, the number $\nt$ of internal events
(figured by crosses in figure~\ref{fig:nested})
grows linearly with time and has relatively negligible fluctuations around its mean value.
In regions~C and~D, $\nt$ grows subextensively in time and keeps fluctuating.
The notation $\th_{1,2}=\infty$ refers to thin-tailed distributions that
possess finite moments of all orders.
Poissonian resetting (see section~\ref{sec:PoissonReset}) lies on the line $\th_2=\infty$.
}
\label{fig:phase}
\end{center}
\end{figure}

The main objective of this study is the statistical analysis of the total number
of internal events (represented by crosses in figure~\ref{fig:nested}) up to time~$t$,
denoted by~$\nt$.
Due to the interplay of two probability distributions, this quantity, despite its apparent simplicity, displays a diverse range of behaviours. 
These are summarised in figure~\ref{fig:phase}, which illustrates the phase diagram of the model in the $\th_1$--$\th_2$-plane.
In this representation, the symbols $\th_{1,2}=\infty$ correspond to thin-tailed
distributions with finite moments of all orders.
This diagram is divided into four distinct regions, exhibiting different
asymptotic forms for the statistics of $\nt$ in the long-time regime of interest.

A remarkable feature of the model emerges from this study.
Specifically, we observe that the more regular of the two renewal processes,
i.e., the one with the larger of the two exponents $\th_1$ and $\th_2$,
always governs the overall regularity of the entire process, as we shall now elaborate.

\begin{enumerate}
\item In regions~A and~B, where the larger exponent is greater than $1$, $\mean{\nt}$ grows linearly in time,
whereas the fluctuations of $\nt$ around its mean value are relatively negligible,
as in the usual framework of renewal theory in the stationary regime ($\th>1$).

More precisely, in region~A, such that $\th_2>1$, we have
\be
\mean{\nt}\approx\frac{\mean{N_{\T}}}{\mean{\T}}\,t,
\ee
where $\mean{N_{\T}}$ is the mean number of internal events between two
resettings, defined in~(\ref{eq:NCox}).
In region~B, such that $\th_2<1<\th_1$,
\be
\mean{\nt}\approx\frac{t}{\mean{\ttau}},
\ee
with the interpretation that, in this region of the phase diagram,
asymptotically, the internal renewal process is not influenced by the external one.

\item In regions~C and~D, where the larger exponent is less than $1$, $\nt$ grows subextensively in time,
with an exponent smaller than unity, and keeps fluctuating,
as in the usual framework of renewal theory in the self-similar regime ($\th<1$).

More precisely, in region~C, where $\th_1<\th_2$,
$\nt$ grows as $t^{\th_2}$ and is asymptotically proportional to the rescaled random
variable $\Xtwo$, which is part of usual renewal theory, and whose density $f_{\Xtwo}(x)$ is
given by~(\ref{eq:fX}),
\be
\nt\approx\mean{N_{\T}}\,\frac{\Xtwo}{\Gamma(1-\th_2)}\left(\frac{t}{T_0}\right)^{\th_2}.
\ee
Thus, $\nt$ is asymptotically equal to the product of the mean number $\mean{N_{\T}}$
of internal events between two resettings by the random number of resettings
(see~(\ref{eq:Ntscal})).

In region~D, where $\th_1>\th_2$,
$\nt$ grows as $t^{\th_1}$ and is asymptotically proportional to a novel rescaled random variable $\Y$,
\be
\nt\approx\frac{\Y}{\Gamma(1-\th_1)}\left(\frac{t}{\tau_0}\right)^{\th_1}.
\ee
The distribution of this dimensionless random variable is universal, depending only on the two exponents $\th_1$ and $\th_2$.
Its probability density is depicted in figure~\ref{fig:half},
for the particular example of $\th_1=1/2$, and for several values of $\th_2$.
\end{enumerate}

This dominance of the more regular process also manifests itself in the two
special cases where either the internal renewal process, or the external one,
are Poisson processes.
In both cases, regardless of the distribution of the other process,
$\nt$ exhibits linear growth over time.
These two situations are different, though.
If the internal process is Poisson, then $\nt$ is Poisson too, regardless of
the distribution of the other process.
If the external (resetting) process is Poisson, the statistics of $\nt$
is exactly described by a single renewal process defined by a
dressed density $\rho^{(\tt r)}(\tau)$, whose expression is given in~(\ref{eq:rhors}),
in terms of the resetting rate $r$
and of the probability density $\rho(\tau)$ of the internal process.
This dressed density is exponentially decaying, regardless of the nature of $\rho(\tau)$.
The superscript in $\rho^{(\tt r)}(\tau)$ is an abbreviation for \textit{replication} or
\textit{resetting}, terms that we shall use interchangeably.

Beyond the analysis of the statistics of $\nt$, a second objective of the present paper is to 
extend the study to other facets of the model and highlight how the ramifications of the theory connect with other studies.
We shall thus be led to consider the question of first-passage time under restart for the process at hand, then revisit some questions related to the study of continuous time random walks under resetting.

The paper is structured as follows.
Section~\ref{sec:over} provides an overview of key concepts and results
in renewal theory that will be used in the subsequent parts of this paper.
The material presented there is classical, with the exception of some more specific results.
Section~\ref{sec:nested} gives the precise definition of the process under study,
as well as the derivation of the key equation~(\ref{eq:key})
for the statistics of $\nt$, which is at the basis of subsequent developments.
Section~\ref{sec:diagram} contains a detailed description of the phase diagram
of the model, summarised above, including all phase boundaries.
Section~\ref{sec:regionD} is devoted to the analysis of the asymptotic
distribution of $\nt$ in region~D ($0<\th_2<\th_1<1$), and to an in-depth study of the universal distribution of the scaling variable $\Y$.
Section~\ref{sec:special} applies the previous formalism to two special cases,
where either the internal process, or the external one, are Poissonian.
Section~\ref{sec:first} deals with the distribution of the first-passage time for the occurrence of a cross (renewal event of the replicated process) in the general case of an arbitrary density $f(T)$, where there is no renewal
description of the sequences of crosses.
Section~\ref{sec:NBt} is focussed on the number
of internal events in the last interval $B_t$ (see figure~\ref{fig:nested}), which is one of
the primary quantities analysed in~\cite{ansell}, thus extending the scope of the
analysis made in this reference to the entire $\th_1$--$\th_2$-plane.
This section also makes the connection of the process under study with continuous time random walks under resetting.

\section{Overview of key concepts in renewal theory}
\label{sec:over}

This section provides an overview of key concepts and results
in renewal theory that will be used in the subsequent parts of this paper.
Classical treatments of the subject can be found
in~\cite{feller1,feller2,cox,cox-miller,grimmett}.
Here we follow the approach presented in~\cite{glrenew} and supplement it with
some additional material.

\subsection{Definition of an ordinary renewal process}

Let us consider events occurring at the random epochs of time $t_1,t_2,\ldots$,
from some time origin $t=0$.
The origin of time is taken on one of these events.
When the intervals of time between events,
$\ttau_1=t_1$, $\ttau_2=t_2-t_1,\ldots$,
are iid random variables with common density $\rho(\tau)$,
the process thus defined is a \textit{renewal process}\footnote{We denote
the random intervals of time $\ttau_1,\ttau_2,\dots$ by bold letters,
and their values in a given realisation of the process by the regular letters
$\tau_1,\tau_2,\dots$.
The same convention applies to the sequence of time
intervals $\T_1,\T_2,\dots$ defined in section~\ref{sec:nested}.
This avoids any ambiguity
(see, e.g., the comment below~(\ref{eq:Ntot})).}.
Otherwise stated, $\ttau_2,\ttau_3,\dots$ are probabilistic copies of the first
time interval $\ttau_1$\footnote{We do not consider here other cases of renewal
processes where the first time interval $\ttau_1$ has a different distribution
from that of the following time intervals $\ttau_2,\ttau_3,\dots$.}.
Hereafter we shall use the terms \textit{event} or \textit{renewal} interchangeably.

A simple example of a renewal process in discrete time is given by
the times of return to the origin of the P\'olya walk mentioned earlier.
An even simpler example arises when considering a continuous time random walk (\textsc{ctrw})~\cite{montroll1965,weiss1994}.
A \textsc{ctrw} is a random walk subordinated to a renewal process.
This means that the waiting times between jumps of the walk, are, by definition, the time intervals of a renewal process.
The jumps are iid random variables $\eta_1, \eta_2, \dots$, with a distribution independent of that of the waiting times. 
In the framework of renewal theory a \textsc{ctrw} is a renewal process with reward~\cite{cox}.
The cumulative process considered in~\cite{ansell}, recalled above, gives an illustration of a \textsc{ctrw}, where the shocks, causing damages of magnitude $\eta_1,\eta_2,\dots$, correspond to the jumps.
The cumulative damage corresponds to the position of the walker.
Moreover, as discussed later (see section~\ref{sec:NBt}), this process is subject to resettings (replacements in~\cite{ansell}).

The survival probability, that is, the probability that no event occurred up to time~$t$
(without counting the event at the origin), is given by
\beq\label{eq:Rt}
R(\tau)=\prob(\ttau >\tau)=\int_\tau^{\infty}\dd t\,\rho(t).
\eeq
The tail behaviour of this distribution plays a crucial role in the subsequent analysis
(and more generally in the study of renewal processes).
It induces a distinction between two main classes of distributions, as summarised below.

\subsubsection*{Thin-tailed distributions}

If the density $\rho(\tau)$ is either supported by a finite interval,
or decaying faster than any power law,
all the moments of the random variable $\ttau$ are finite.
The Laplace transform of $\rho(\tau)$,
where $s$ is conjugate to $\tau$,
is then given by the power series
\be
\lap{\tau}\rho(\tau)=
\hat{\rho}(s)=\mean{\e^{-s\ttau}}=\sum_{k\ge0}\frac{(-s)^k}{k!}\mean{\ttau^k}.
\ee
More specifically,
the above series is convergent if $\rho(\tau)$ either has finite support
or decays exponentially or faster,
whereas it is only a formal power series if the decay of $\rho(\tau)$
is slower than exponential.

\subsubsection*{Fat-tailed distributions}

If $\rho(\tau)$ is characterised by a power-law fall-off with an arbitrary
index $\th>0$,
parametrising its tail as
\beq\label{eq:Ras}
R(\tau)\approx
\left(\frac{\tau_0}{\tau}\right)^{\th},
\eeq
where $\tau_0$ is a microscopic time scale,
we have
\beq\label{eq:roas}
\rho(\tau)\approx\frac{c}{\tau^{1+\th}},\qquad c=\th\tau_0^\th.
\eeq
Here, $\ttau$ has only finitely many moments,
as $\mean{\ttau^k}$ is convergent only for $k<\th$.

For any value of the index $\th$ that is not an integer,
the Laplace transform $\hat\rho(s)$ of the density has a singular part as
$s\to0$,
of the form
\be
\hat{\rho}(s)_\sg\approx c\,\Gamma(-\th)s^\th.
\ee
We thus have
\beq\label{eq:broad}
\hat{\rho}(s)\approx\left\{
\begin{array}{ll}
1-a\,s^\th & (\th<1),\\
1-\mean{\ttau}s+a\,s^\th\quad & (1<\th<2),
\end{array}
\right.
\eeq
and so on,
with more regular terms
as $\th$ lies between higher consecutive integers,
and where the positive amplitude $a$ reads
\beq\label{eq:defa}
a=c\,\abs{\Gamma(-\th)}=\abs{\Gamma(1-\th)}\tau_0^\th.
\eeq

Whenever the index $\th$ is an integer,
$\hat\rho(s)$ is affected by logarithmic corrections.
We mention for further reference the case $\th=1$,
where we have
\be
R(\tau)\approx\frac{\tau_0}{\tau},\qquad
\rho(\tau)\approx\frac{\tau_0}{\tau^2}
\ee
and
\beq\label{eq:thone}
\hat\rho(s)\approx1+\tau_0s\ln(\tau_\star s).
\eeq
This expression involves, in general, two different microscopic time scales,
the amplitude $\tau_0$ (describing the tail of the distribution)
and the finite part $\tau_\star$ (depending on details of the whole distribution).

The class of thin-tailed distributions,
where all the moments of $\ttau$ are finite,
corresponds formally to $\th=\infty$.

\subsection{The number $N_t$ of renewals}

The number of renewals $N_t$ that occur in the time interval $(0,t)$ satisfies
the condition
\beq\label{eq:Nttn}
\prob(N_t\ge n)=\prob(t_n\le t),
\eeq
where the sum of the first $n$ time intervals,
\beq\label{eq:tndef}
t_n=\ttau_1+\ttau_2+\cdots+\ttau_n,
\eeq
is the waiting time until the occurrence of the $n$th event, or, for short, the
time of the $n$th renewal.
Correspondingly, the time intervals $\ttau_1,\ttau_2,\dots$ obey the sum
rule
\beq\label{eq:sum}
\ttau_1+\ttau_2+\cdots+\ttau_{N_t}+b_t=t,
\eeq
where $b_t$ is the backward recurrence time,
or the age of the renewal process at time $t$,
which measures the time elapsed since the last renewal event.
The distribution of $N_t$ is given by
\beqa\label{eq:Nt}
p_n(t)&=&\prob(N_t=n)
\nonumber\\
&=&\int\dd\tau_1\dots\dd\tau_n\,\dd b\,\rho(\tau_1)\dots\rho(\tau_n)R(b)
\,\delta\Big(\sum_{i=1}^n\tau_i+b-t\Big)
\nonumber\\
&=&((\rho\star)^n\star R)(t),
\eeqa
where the star denotes a temporal convolution and $(\rho\star)^n$ denotes the
$n$th convolution of the density $\rho(t)$.
We have in particular
\beq\label{eq:pzero}
p_0(t)=R(t)
\eeq
(see~(\ref{eq:Rt})).
In Laplace space,~(\ref{eq:Nt}) reads
\beq\label{eq:pnlap}
\hat p_n(s)=\lap{t}p_n(t)=\hat\rho(s)^n\,\hat R(s),
\eeq
with
\beq\label{eq:Rs}
\hat R(s)=\frac{1-\hat\rho(s)}{s}.
\eeq

The distribution of $N_t$ can be expressed compactly through its probability
generating function
\beq\label{eq:ZNt}
Z(z,t)=\mean{z^{N_t}}=\sum_{n\ge0}z^n\,p_n(t).
\eeq
Using~(\ref{eq:pnlap}),~(\ref{eq:Rs}), this yields, in Laplace space,
\be
\hat Z(z,s)=\lap{t}\mean{z^{N_t}}
=\sum_{n\ge0}\,z^n\,\hat p_n(s)
=\frac{\hat R(s)}{1-z\hat\rho(s)},
\ee
i.e.,
\beq\label{eq:Zzs}
\hat Z(z,s)=\frac{1-\hat\rho(s)}{s(1-z\hat\rho(s))}.
\eeq
Note that $\hat Z(1,s)=1/s$, as it should be.
Expressions for the moments $\mean{N_t^k}$ in Laplace space can be obtained by
differentiating~(\ref{eq:Zzs}) with respect to $z$ at $z=1$.
We obtain in particular
\beqa
\lap{t}\mean{N_t}
&=&z\frac{\partial}{\partial z}\hat Z(z,s)\Big\vert_{z=1}
=\frac{\hat\rho(s)}{s(1-\hat\rho(s))},
\label{eq:Nt1}
\\
\lap{t}\mean{N_t^2}
&=&\Big(z\frac{\partial}{\partial z}\Big)^2\hat Z(z,s)\Big\vert_{z=1}
=\frac{\hat\rho(s)(1+\hat\rho(s))}{s(1-\hat\rho(s))^2}.
\label{eq:Nt2}
\eeqa

\subsection{Mean of the single-interval distribution}
Another quantity of interest for the sequel is the mean of the single-interval distribution,
that is, the distribution of any of the intervals $\ttau_1,\ttau_2,\dots,\ttau_{N_t}$ subject to the condition~(\ref{eq:sum}).
This observable, denoted by $\ttau_t$ is defined
 provided that $N_t\ge1$.
In the event where $N_t=0$,
which occurs with probability $p_0(t)$ given by~(\ref{eq:pzero}), $\ttau_t$ is
conventionally set to zero and therefore does not contribute to its mean.
We thus have, taking $\ttau_t$ to be the first interval,
\beq\label{eq:deftaut}
\mean{\ttau_t}=\sum_{n\ge0}\int\dd\tau_1\dots\dd\tau_n\,\dd
b\,\tau_1\,\rho(\tau_1)\dots\rho(\tau_n)R(b)
\,\delta\Big(\sum_{i=1}^n\tau_i+b-t\Big).
\eeq
In Laplace space, it is readily found that~\cite{glrenew}
\beq\label{eq:taumeans}
\lap{t}\mean{\ttau_t}
=\frac{1}{s}\int_{0}^{\infty}\dd\tau\,\tau\,\rho (\tau){\e}^{-s\tau }
=-\frac{1}{s}\frac{\dd\hat{\rho}(s)}{\dd s}.
\eeq

\subsection{Asymptotic distribution of the time of the $n$th renewal}
\label{sec:tn}

As can be seen on~(\ref{eq:Nttn}), the two quantities $N_t$ and $t_n$ represent
complementary facets of a renewal process.
In particular the asymptotic behaviours of these quantities in the long-time
regime go hand in hand.
We start by discussing the simpler case of~$t_n$,
before delving in that of $N_t$ in the next section.

As we now show, when the number $n$ of time intervals becomes large,
the asymptotic growth of $t_n$ obeys the following dichotomy,
dictated by the law of large numbers.

\subsubsection*{Finite $\mean{\ttau}$}

In this case, i.e., for $\th>1$,
we have
\beq\label{eq:tave}
\mean{t_n}=n\mean{\ttau}.
\eeq
According to the law of large numbers,
$t_n/n\to\mean{\ttau}$ when $n\to\infty$, in probability.
This essentially means that typical fluctuations of $t_n$ around its mean value
grow less rapidly than linearly in $n$.
These fluctuations are therefore subextensive, i.e., relatively negligible.
They can be characterised more precisely as follows.
 If $\var{\ttau}=\mean{\ttau^2}-\mean{\ttau}^2$ is finite, i.e., for
$\th>2$,
then
\be
\var t_n=n\var{\ttau}.
\ee
According to the central limit theorem,
the difference $t_n-n\mean{\ttau}$ grows as $\sqrt{n}$,
and has an asymptotic normal distribution.
If, on the other hand, $\var{\ttau}$ is divergent, i.e., for $1<\th<2$,
the difference $t_n-n\mean{\ttau}$ grows as $n^{1/\th}$,
and its asymptotic distribution is a L\'evy stable law.

\subsubsection*{Divergent $\mean{\ttau}$}

In this case, i.e., for $\th<1$,
the law of large numbers does not apply.
The sum $t_n$ grows more rapidly than linearly in $n$ and keeps fluctuating.
Using~(\ref{eq:broad}) and (\ref{eq:tndef}),
we have indeed
\be
\mean{\e^{-st_n}}=\hat\rho(s)^n\approx\e^{-n a s^\th},
\ee
and therefore (see~(\ref{eq:defa}))
\beq\label{eq:tell}
t_n\approx
(an)^{1/\th}\,L_\th=\tau_0\left(\Gamma(1-\th)\,n\right)^{1/\th}\,L_\th,
\eeq
where the rescaled random variable $L_\th$
is distributed according to the normalised one-sided L\'evy stable law of index $\th$,
with density
\beq\label{eq:fth}
f_{L_\th}(x)=\int\frac{\dd s}{2\pi\ii}\,\e^{sx-s^\th}\qquad(0<x<+\infty).
\eeq
The power-law tail
\be
f_{L_\th}(x)\approx\frac{\th}{\Gamma(1-\th)\,x^{1+\th}}
\ee
mirrors that of the underlying density $\rho(\tau)$.

\subsubsection*{Marginal situation}

When $\th=1$,
the first moment $\mean{\ttau}$ diverges logarithmically,
thus the law of large numbers is affected by logarithmic corrections.
Using~(\ref{eq:thone}),
we have indeed
\be
\mean{\e^{-st_n}}\approx\e^{n\tau_0s\ln(\tau_\star s)},
\ee
and therefore
\beq\label{eq:tmargin}
t_n\approx n\tau_0\left(\ln\frac{n\tau_0}{\tau_\star}+\Xi\right).
\eeq
The expression between the parentheses is the sum of a deterministic component,
which grows logarithmically with $n$,
and of a finite fluctuating part $\Xi$,
following a Landau distribution~\cite{landau},
\beq\label{eq:landau}
f_\Xi(\xi)=\int\frac{\dd z}{2\pi\ii}\,\e^{z\xi+z\ln
z}\qquad(-\infty<\xi<+\infty).
\eeq
The right tail of this distribution,
\be
f_\Xi(\xi)\approx\frac{1}{\xi^2}\qquad(\xi\to+\infty),
\ee
mirrors that of the underlying density $\rho(\tau)$,
and implies that $\mean{\Xi}$ is logarithmically divergent.

\subsection{Asymptotic distribution of the number $N_t$ of renewals}
\label{sec:Ntas}

The results summarised below highlight the close connection
between the statistics of the sum $t_n$ of the first $n$ time intervals
for a fixed large number $n$ of intervals,
and of the number $N_t$ of renewals
up to a fixed large observation time~$t$.
In particular, the dichotomy between finite and infinite mean
$\mean{\ttau}$, described in section~\ref{sec:tn}, also prevails for the
asymptotic distribution of $N_t$ (see, e.g.,~\cite{feller2,glrenew}).

\subsubsection*{Finite $\mean{\ttau}$}

In this case, i.e., for $\th>1$, the analysis of~(\ref{eq:Nt1})
and~(\ref{eq:Nt2}) shows that the mean number of events scales as
\beq\label{eq:Ntnarrow}
\mean{N_t}\approx\frac{t}{\mean{\ttau}},
\eeq
and $N_t$ exhibits subextensive,
i.e., relatively negligible, fluctuations around this mean value.
The two ensembles defined above are therefore equivalent,
in the sense used in thermodynamics.
In other words,
time and the number of events are asymptotically proportional to each other,
as testified by~(\ref{eq:tave}) and~(\ref{eq:Ntnarrow}).
Furthermore, these quantities are tightly related,
in the sense that the relative fluctuations between the two quantities are negligible.

The fluctuations of $N_t$ can be characterised more precisely as follows.
If $\var{\ttau}$ is finite, i.e., for $\th>2$, we have
\be
\var N_t\approx\frac{\var{\ttau}}{\mean{\ttau}^3}\,t.
\ee
The difference $N_t-t/\mean{\ttau}$ grows as $\sqrt{t}$,
and has an asymptotic normal distribution.
If $\var{\ttau}$ is divergent, i.e., for $1<\th<2$,
the difference $N_t-t/\mean{\ttau}$ grows as $t^{1/\th}$
and its asymptotic distribution is a L\'evy stable law.
To mention a further subtlety,
the variance of $N_t$ grows as $t^{3-\th}$,
with an exponent larger than the exponent $2/\th$
describing typical square fluctuations.
The exponent difference $3-\th-2/\th=(2-\th)(\th-1)/\th$
is positive and vanishes both for $\th\to1$ and for $\th\to2$~\cite{glrenew}.

Renewal processes with a finite
mean time interval $\mean{\ttau}$ become stationary in the regime of late times.
One-time observables,
such as the distribution of $B_t$ or of the excess time $E_t=t_{N_t+1}-t$,
reach well-defined limiting forms,
whereas two-time quantities depend solely on the difference between the two times, asymptotically.
For instance, the number of renewals between times $t$ and $t+t'$ only depends on the time separation $t'$, when $t$ is large~\cite{glrenew}.

\subsubsection*{Divergent $\mean{\ttau}$}
In this case, i.e., for $\th<1$,
the number $N_t$ of events grows sublinearly with time and keeps fluctuating.
Its mean value can be obtained from~(\ref{eq:broad}),~(\ref{eq:Nt1}), yielding
\beq\label{eq:Ntrenew}
\mean{N_t}\approx
\frac{\sin\pi\th}{\pi\th}\,\left(\frac{t}{\tau_0}\right)^\th.
\eeq
Moreover, its full distribution can be derived from~(\ref{eq:pnlap}), which translates to
\beq\label{eq:plap}
\hat p_n(s)\approx as^{\th-1}\,\e^{-na s^\th}.
\eeq
Thus, setting
\beq\label{eq:Ntscal}
N_t\approx\X\,\frac{t^\th}{a}=
\frac{\X}{\Gamma(1-\th)}\left(\frac{t}{\tau_0}\right)^\th,
\eeq
we find that the probability density of the rescaled random variable $\X$ is
\beq\label{eq:fX}
f_{\X}(x)=\int\frac{\dd z}{2\pi\ii}\,z^{\th-1}\e^{z-xz^\th}\qquad (x>0),
\eeq
which entails that the random variable $\X$ can be written
as~\cite{pollard,planaires}
\beq\label{eq:xleq}
\X=L_\th^{-\th},
\eeq
where the distribution of $L_\th$
is the L\'evy stable law~(\ref{eq:fth}).
This manifests the equivalence of~(\ref{eq:tell}) and (\ref{eq:Ntscal}), under
the replacement of $N_t$ by $n$, and $t_n$ by $t$:
in the two ensembles defined above, the number of events scales as a power of
time with exponent $\th$.
However, at variance with the situation where $\mean{\ttau}$ is finite,
the asymptotic relations~(\ref{eq:tell}) and (\ref{eq:Ntscal}) between time and
the number of events
involve a fluctuating variable, denoted by $L_\th$ or $\X$, respectively.

Renewal processes with a divergent $\mean{\ttau}$
exhibit self-similar behaviour and universality at late times.
In particular,
dimensionless observables such as the ratios $B_t/t$ and $E_t/t$,
as well as the rescaled occupation time,
have non-trivial limiting distributions,
that depend solely on the exponent $\th$.
Furthermore, the non-stationarity of the processes imply that two-time quantities depend on both instances of time, asymptotically.
For instance, the number of renewals between times $t$ and $t+t'$ now depends on both the waiting time $t$ and the time separation $t'$, a property referred to as aging~\cite{glrenew}.

We also provide, for future reference, several results pertaining to
the density $f_{\X}(x)$ of the rescaled random variable $\X$.
The integral representation~(\ref{eq:fX})
implies that $f_{\X}(x)$ is regular at small $x$:
\beq\label{eq:fsmall}
f_{\X}(x)=\frac{1}{\Gamma(1-\th)}-\frac{x}{\Gamma(1-2\th)}+\cdots
\eeq
A saddle-point treatment shows that
this density decays as a compressed exponential at large $x$, for all
$0<\th<1$, according~to
\beq\label{eq:flarge}
f_{\X}(x)\sim\exp\left(-(1-\th)(\th^\th x)^{1/(1-\th)}\right).
\eeq
This fast decay has two consequences.
First, all the moments of $\X$ are finite.
They are given by the explicit formula
\beq\label{eq:momentX}
\mean{\X^k}=\frac{k!}{\Gamma(k\th+1)}.
\eeq
Second, the corresponding Laplace transform $\hat f_{\X}(u)$
is an entire function in the whole $u$-plane.
This reads explicitly~\cite{feller2,pollard}
\beq\label{eq:mittag}
\hat f_{\X}(u)
=\int_0^\infty\dd x\,\e^{-ux}f_{\X}(x)
=\sum_{k\ge0}\frac{(-u)^k}{\Gamma(k\theta+1)}=\mathrm{E}_\theta(-u),
\eeq
where $\mathrm{E}_\th(z)$ is the Mittag-Leffler function
of index $\th$ (see~\cite{HMS} for a review)\footnote{
The distribution
of the random variable $X_\th$ is named Mittag-Leffler
by some authors~\cite{darling}.
Another definition of the Mittag-Leffler distribution is used in other works, though
(see, e.g.,~\cite{pillai}).}.

\begin{figure}
\begin{center}
\includegraphics[angle=0,width=.75\linewidth,clip=true]{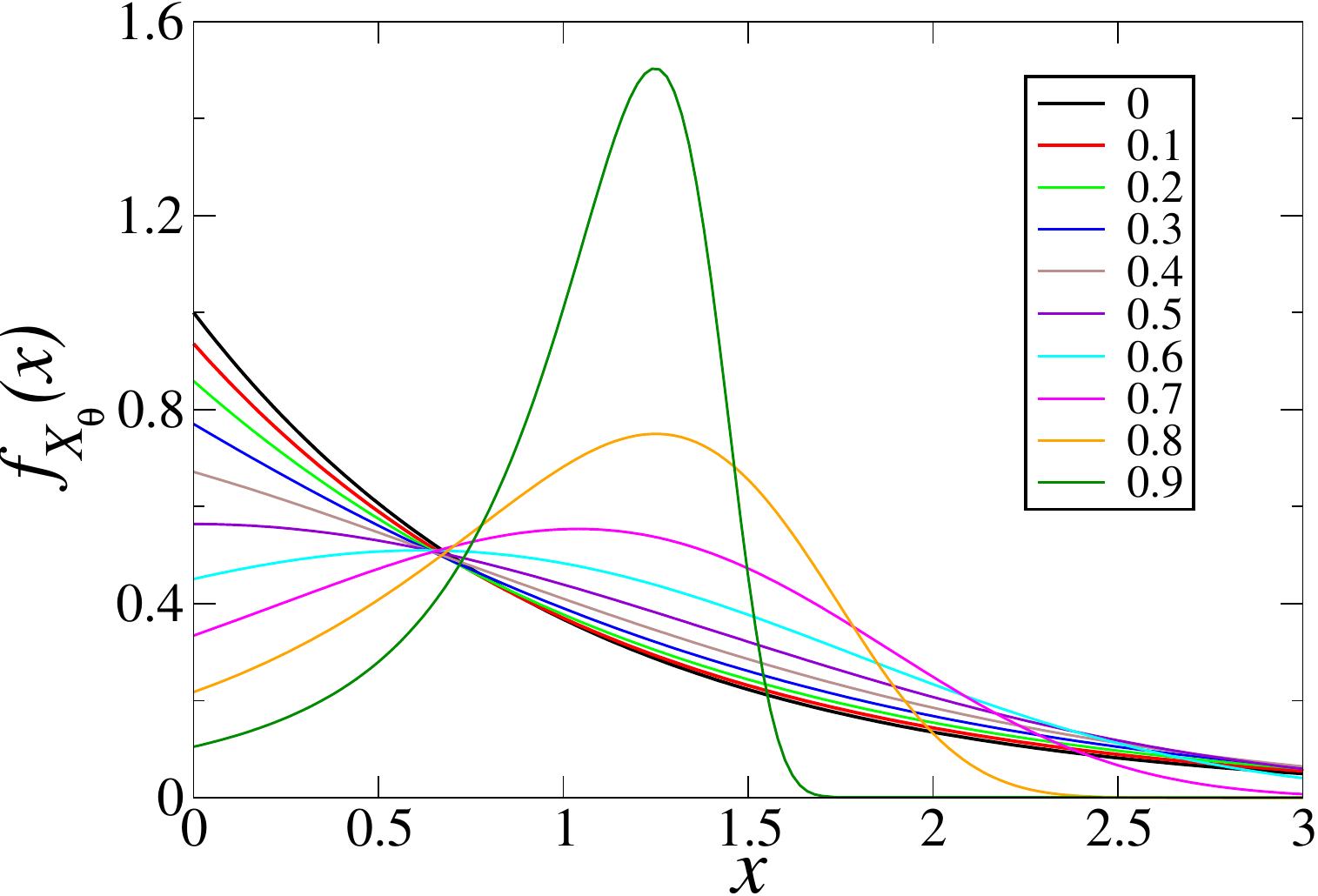}
\caption{\small
Probability density $f_{\X}(x)$ of the rescaled random variable $\X$
entering~(\ref{eq:Ntscal}),
for several values of the index $\th$ (see legend).}
\label{fig:levy}
\end{center}
\end{figure}

Figure~\ref{fig:levy} shows plots of the density $f_{\X}(x)$ for several values
of the index $\th$ (see legend).
This distribution is a monotonically decreasing function of $x$ for $\th<1/2$,
whereas it exhibits a non-trivial maximum for $1/2<\th<1$.

For $\th\to0$, the distribution of $\X$ becomes a simple exponential, with density
\be
f_{X_0}(x)=\e^{-x},\qquad\hat f_{X_0}(u)=\frac{1}{1+u}.
\ee

For $\th=1/2$, the distribution of $\X$ is a half-Gaussian, with density
\beq\label{eq:12}
f_{X_{1/2}}(x)=\frac{\e^{-x^2/4}}{\sqrt\pi},\qquad\hat
f_{X_{1/2}}(u)=\e^{u^2}\erfc u,
\eeq
where $\erfc$ is the complementary error function.

For $\th\to1$, the distribution of $\X$ becomes degenerate:
\beq\label{eq:1}
f_{X_{1}}(x)=\delta(x-1),\qquad\hat f_{X_{1}}(u)=\e^{-u}.
\eeq

\subsubsection*{Marginal situation}

When $\th=1$,
the first moment $\mean{\ttau}$ diverges logarithmically
and the above results are affected by logarithmic corrections.
Let us focus on the mean number $N_t$ of events.
Inserting the asymptotic expression~(\ref{eq:thone}) of $\hat\rho(s)$
into~(\ref{eq:Nt1}),
we obtain
\be
\lap{t}\mean{N_t}\approx-\frac{1}{\tau_0s^2\ln(\tau_\star s)},
\ee
hence
\beq\label{eq:nmargin}
\mean{N_t}\approx\frac{1}{\ln(t/\tau_\star)+\gamma-1}\,\frac{t}{\tau_0},
\eeq
where $\gamma$ is Euler's constant.

Typical fluctuations of $N_t$ around its mean value are again relatively negligible,
albeit marginally,
as their typical size is smaller than $\mean{N_t}$ by one power of
$\ln(t/\tau_\star)$.
Skipping details, let us mention the formula
\beq\label{eq:nfullmargin}
N_t\approx\frac{1}{\ln(t/\tau_\star)+\Xi}\,\frac{t}{\tau_0},
\eeq
where the random variable $\Xi$ has the Landau distribution~(\ref{eq:landau}).
A comparison between~(\ref{eq:tmargin}) and~(\ref{eq:nfullmargin})
again demonstrates a tight match between the two aforementioned ensembles.
Note however that there is no simple connection between~(\ref{eq:nmargin})
and~(\ref{eq:nfullmargin}), as $\mean{\Xi}$ is divergent.

\subsection{Asymptotic behaviour of the mean of the single-interval distribution}

The asymptotic analysis of the quantity defined in~(\ref{eq:deftaut}) can be
done along the same lines as above, using~(\ref{eq:taumeans}), and leads to the following results.

If, first, $\mean{\ttau}$ is finite, i.e., $\th>1$,
then $\mean{\ttau_t}$ converges to $\mean{\ttau}$,
in line with~(\ref{eq:Ntnarrow}).
On the other hand, if $\mean{\ttau}$ is divergent, i.e., $\th<1$, we have
\beq\label{eq:meantauR}
\mean{\ttau_t}\approx\frac{\th\tau_{0}^{\th}}{1-\th}t^{1-\th}.
\eeq
The product of this quantity by the mean number of events (see~(\ref{eq:Ntrenew})),
\be
\mean{N_t}\mean{\ttau_t}\approx\frac{\sin\pi\th}{\pi(1-\th)}\,t,
\ee
grows linearly in time,
with a universal amplitude depending only on the exponent $\th$.
In the marginal situation where $\th=1$,
substituting~(\ref{eq:thone}) in~(\ref{eq:taumeans}) yields
\be
\lap{t}\mean{\ttau_t}\approx-\frac{\tau_0}{s}(\ln(\tau_\star s)+1),
\ee
hence
\beq\label{eq:singlemean}
\mean{\ttau_t}\approx\tau_0\Big(\ln\frac{t}{\tau_\star}+\gamma-1\Big).
\eeq
Thus, interestingly, (\ref{eq:nmargin}) can be rewritten as
\be
\mean{N_t}\approx\frac{t}{\mean{\ttau_t}}.
\ee
More precisely,
\be
\mean{N_t}\mean{\ttau_t}=\left(1+O\Big(\frac{1}{(\ln t)^2}\Big)\right)t.
\ee

\section{Two nested renewal processes: definition and general results}
\label{sec:nested}

As depicted in figure~\ref{fig:nested}, the stochastic process under study is
obtained by the
replication of a renewal process, defined by the sequence
$\ttau_1,\ttau_2,\dots$
of iid random intervals of time, with common probability
density~$\rho(\tau)$---\textit{the internal renewal process}---according to
another renewal process, defined by the sequence $\T_1,\T_2,\dots$
of iid random intervals of time, with common probability density
$f(T)$---\textit{the external renewal process}.
Events separated by the time intervals $\ttau_1,\ttau_2,\dots$,
shown as crosses in figure~\ref{fig:nested},
are referred to as internal events,
whereas events separated by the time intervals $\T_1,\T_2,\dots$, shown as dots,
are referred to as external events.
We shall alternatively refer to external events as resetting events, since
the process defined in this manner can also be interpreted as a renewal
process, characterised by the density $\rho(\tau)$, that is reset at random time
intervals, drawn from the density~$f(T)$.

In the following, our objective is to analyse the stochastic process made of
these two nested renewal processes,
with a focus on the statistics of the number $\nt$ of internal events occurring
up to time $t$.
We shall be mostly interested in the case where the two probability densities
of the internal and external processes have power-law decays
of the form
\beq\label{eq:t1t2}
\rho(\tau)\approx\frac{\th_1\,\tau_0^{\th_1}}{\tau^{1+\th_1}},\qquad
f(T)\approx\frac{\th_2\,T_0^{\th_2}}{T^{1+\th_2}},
\eeq
with arbitrary positive exponents $\th_1$, $\th_2$.
As mentioned earlier, thin-tailed distributions with finite moments of all orders
formally correspond to taking infinite values for these exponents.

The number of resetting events,
i.e., of time intervals $\T_1,\T_2,\dots$, up to time $t$ is denoted by $M_t$.
These intervals obey the sum rule
\beq\label{eq:contrainte}
\T_1+\T_2+\cdots+\T_{M_t}+B_t=t,
\eeq
where the backward recurrence time $B_t$ is, as previously defined, the time
elapsed
since the last resetting event.
Within each interval $\T_i$, there are $N_{\T_i}$ internal events induced by
$\rho(\tau)$.
The total number of these internal events up to time $t$ is given by
\beq\label{eq:Ntot}
\nt =N_{\T_1}+N_{\T_2}+\cdots+N_{\T_{M_t}}+N_{B_t}.
\eeq
In this expression, $N_{\T_1},N_{\T_2},\dots$ are doubly stochastic quantities,
to be distinguished from $N_{T_1},N_{T_2},\dots$,
since the time variables $\T_1,\T_2,\dots$ are random variables themselves.
The former are averaged over both internal and external processes, the latter
on the internal process only (see~(\ref{eq:NCox})).

All information on the distribution of $\nt$ is encoded in the generating
function
\be
\Z(z,t)=\mean{z^{\nt}},
\ee
where the average is taken over the realisations $\cc=\{T_1,T_2,\dots,B\}$ of
the external variables $\T_1,\T_2,\dots,B_t$,
with the weight (for fixed $M_t=m$)
\beq\label{eq:Pcc}
P(\cc)=f(T_1)\dots f(T_m)\Phi(B)\,\delta\Big(\sum_{i=1}^{m} T_i+B-t\Big),
\eeq
where
\be
\Phi(B)=\prob(\T>B)=\int_B^\infty\dd T\,f(T)
\ee
is the survival probability of the external process,
and over the realisations $\tilde\cc=\{\tau_1,\tau_2,\dots,b\}$ of the internal
variables, $\ttau_1,\ttau_2,\dots,b_T$,
attached to each interval $\T_1,\T_2,\dots$,
with the weight (for fixed $N_T=n$)
\be
P(\tilde\cc)=
\rho(\tau_1)\dots\rho(\tau_n)R(b)
\,\delta\Big(\sum_{i=1}^n\tau_i+b-T\Big).
\ee
Thus
\be
\Z(z,t)=\sum_{\cc}P(\cc)\sum_{\tilde
\cc_1,\tilde\cc_2,\dots}P(\tilde\cc_1)z^{N_{T_1}}P(\tilde\cc_2)z^{N_{T_2}}\dots,
\ee
with the notations
\be
\sum_{\cc}=\sum_{m\ge0}\int_0^\infty\dd T_1\dots\dd T_m\,\dd B,
\quad\sum_{\tilde\cc}=\sum_{n\ge0}\int_0^\infty\dd\tau_1\dots\dd\tau_n\,\dd b.
\ee
The average over the internal variables of each term $z^{N_{T_i}}$ with the
weight
$P(\tilde\cc)$ gives a factor $Z(z,T_i)$ (see~(\ref{eq:ZNt})).
We then average over the external variables with the weight $P(\cc)$ to obtain
\beq\label{eq:calN}
\Z(z,t)=\sum_{\cc}P(\cc)Z(z,T_1)\dots Z(z,T_m)Z(z,B).
\eeq
This expression is a convolution,
which is easier to handle in Laplace space, leading to
\beq\label{eq:hatZcal}
\hat\Z(z,s)=\lap{t}\Z(z,t)
=\sum_{m\ge0}
\int_0^\infty\dd T_1\,\e^{-sT_1}f(T_1)Z(z,T_1)
\dots
\int_0^\infty\dd B\,\e^{-sB}\Phi(B)Z(z,B)
\eeq
with
\beqa
\hat\phi(z,s)
&=&\int_0^\infty\dd T\,\e^{-sT}f(T)Z(z,T),
\label{eq:phidef}
\\
\hat\psi(z,s)
&=&\int_0^\infty\dd T\,\e^{-sT}\Phi(T)Z(z,T),
\label{eq:psidef}
\eeqa
thus finally
\beq\label{eq:key}
\hat\Z(z,s)
=\frac{\hat\psi(z,s)}{1-\hat\phi(z,s)}.
\eeq
This key equation is the starting point of all forthcoming developments.

Expressions for the moments $\mean{\nt^k}$ in Laplace space
can be obtained by differentiating~(\ref{eq:key}) with respect to $z$ at $z=1$,
along the lines of~(\ref{eq:Nt1}).
We thus obtain
\beqa
\lap{t}\mean{\nt}
&=&\frac{I_1(s)}{s(1-\hat f(s))}+\frac{I_2(s)}{1-\hat f(s)},
\label{eq:nc1}
\\
\lap{t}\mean{\nt^2}
&=&\frac{I_1(s)+sI_2(s)+I_3(s)+sI_4(s)}{s(1-\hat f(s))}
+\frac{2I_1(s)(I_1(s)+sI_2(s))}{s(1-\hat f(s))^2},
\label{eq:nc2}
\eeqa
with the definitions
\beqa
I_1(s)&=&\int_0^\infty\dd T\,\e^{-sT}f(T)\mean{N_T},\quad
I_2(s)=\int_0^\infty\dd T\,\e^{-sT}\Phi(T)\mean{N_T},
\nonumber\\
I_3(s)&=&\int_0^\infty\dd T\,\e^{-sT}f(T)\mean{N_T^2},\quad
I_4(s)=\int_0^\infty\dd T\,\e^{-sT}\Phi(T)\mean{N_T^2}.
\label{eq:II}
\eeqa

Let us note, for later reference (see section~\ref{sec:NBt}), that, by
applying the same reasoning, we can
obtain the expression of $\mean{z^{N_{B_t}}}$ in Laplace space.
Indeed, we have (see~(\ref{eq:calN}))
\be
\mean{z^{N_{B_t}}}=\sum_{\cc}P(\cc)Z(z,B),
\ee
hence (see~(\ref{eq:hatZcal}))
\beq\label{eq:B}
\lap{t}\mean{z^{N_{B_t}}}
=\frac{\hat\psi(z,s)}{1-\hat f(s)}.
\eeq
It follows that
\beq\label{eq:probNB}
\lap{t}\prob(N_{B_t}=n)=\frac{\int_0^\infty\dd T\,\e^{-sT}\Phi(T)p_n(T)}{1-\hat
f(s)},
\eeq
and therefore
\beq\label{eq:probNBk}
\lap{t}\mean{N_{B_t}^k}=\frac{\int_0^\infty\dd
T\,\e^{-sT}\Phi(T)\mean{N_T^k}}{1-\hat f(s)}.
\eeq
In particular,
\beq\label{eq:NB}
\lap{t}\mean{N_{B_t}}=\frac{I_2(s)}{1-\hat f(s)},
\eeq
which is the second term on the right-hand side of~(\ref{eq:nc1}).
Accordingly, the first term on the right-hand side of the latter equation represents
the Laplace transform of the mean sum of the first $M_t$ terms in~(\ref{eq:Ntot}).

\section{Phase diagram}
\label{sec:diagram}

The asymptotic behaviour of the number $\nt$ of internal events in the long-time regime
is determined by the characteristics
of the underlying probability densities~$\rho(\tau)$ and~$f(T)$,
and chiefly by their tail exponents $\th_1$ and $\th_2$.
The subsequent analysis evidences four regions,
labelled in order of increasing complexity,
and depicted in the phase diagram presented in figure~\ref{fig:phase}.
The behaviour of $\nt$ along the boundaries between these regions
is considered at the end of this section.

\subsubsection*{Region~A $(\th_2>1)$}

When the first moment $\mean{\T}$ of the external density $f(T)$ is finite,
that is, if $\th_2>1$,
the first term in the right-hand side of~(\ref{eq:nc1}) gives the leading
contribution as $s\to0$ (see~section~\ref{sec:NBt} for an analysis of the second term).
The integral~$I_1(s)$ defined in~(\ref{eq:II}) has a finite limit for $s\to0$,
which represents the mean number $\mean{N_{\T}}$ of internal events
in the random interval $(0,\T)$~\cite{cox,cox60},
i.e.,
\beq\label{eq:NCox}
I_1(0)=\mean{N_{\T}}=\int_0^\infty\dd
T\,f(T)\mean{N_T}=\Big\langle\sum_{n\ge0}n\,p_n(\T)\Big\rangle,
\eeq
where the average in the rightmost expression pertains to the random variable $\T$.
The denominator of the first term in the right-hand side of~(\ref{eq:nc1}) behaves
as $\mean{\T}s^2$, thus, finally,
\beq\label{eq:NregionA}
\mean{\nt}\approx\frac{\mean{N_{\T}}}{\mean{\T}}\,t.
\eeq
The interpretation of this result is intuitively clear:
asymptotically, $\mean{\nt}$ is
the product of the mean number $\mean{N_{\T}}$
of internal events between two resettings
by the mean number of resettings in $(0,t)$ (see~(\ref{eq:Ntnarrow})),
\beq\label{eq:Nnut}
\mean{\nt}
\approx\mean{N_{\T}}\mean{M_t}.
\eeq
Similarly, the square of $I_1(0)$
gives the leading contribution to~(\ref{eq:nc2}) as $s\to0$.
We thus obtain the estimate
\be
\mean{\nt^2}\approx\frac{\mean{N_{\T}}^2}{\mean{\T}^2}t^2
\approx\mean{N_{\T}}^2\mean{M_t}^2,
\ee
demonstrating that typical fluctuations of $\nt$
around its mean value~(\ref{eq:NregionA}) are relatively negligible.

\subsubsection*{Region~B $(\th_2<1<\th_1)$}

In this region, since $\th_2<1$, the first moment of the external density
$f(T)$ is divergent.
In the present context, (\ref{eq:Ras}), (\ref{eq:roas}), (\ref{eq:broad}) and
(\ref{eq:defa}) become
\beqa
f(T)&\approx&\frac{\th_2T_0^{\th_2}}{T^{1+\th_2}},\qquad
\Phi(T)\approx\frac{T_0^{\th_2}}{T^{\th_2}},
\label{eq:th21}
\\
1-\hat f(s)&\approx& a_2s^{\th_2},
\qquad a_2=\Gamma(1-\th_2)T_0^{\th_2}.
\label{eq:th22}
\eeqa
The integrals $I_1(s)$ and $I_2(s)$ are divergent for $s\to0$,
and their leading behaviour can be estimated by using $\mean{N_T}\approx
T/\mean{\ttau}$.
We thus obtain
\be
I_1(s)+sI_2(s)\approx\frac{a_2}{\mean{\ttau}}\,s^{\th_2-1},
\ee
where both integrals contribute to the above estimate,
hence
\be
\lap{t}\mean{\nt}\approx\frac{1}{\mean{\ttau}s^2},
\ee
and finally
\beq\label{eq:region3}
\mean{\nt}\approx\frac{t}{\mean{\ttau}}.
\eeq
The integrals $I_3(s)$ and $I_4(s)$ are also divergent as $s\to0$.
Some algebra yields the estimate
\be
\mean{\nt^2}\approx\frac{t^2}{\mean{\ttau}^2},
\ee
showing that typical fluctuations of $\nt$ around its mean value
are again relatively negligible.
To conclude, asymptotically, the internal process is not influenced by the
external one, as far as the mean $\mean{\nt}$ is concerned.

\subsubsection*{Region~C $(\th_1<\th_2<1)$}

In this region, both exponents are less than unity,
which implies that large fluctuations in the statistics of $\nt$ are to be
expected.
Since $\th_1<\th_2$, it is also expected that the number
of internal events between any two consecutive resettings
is typically finite and of the order of $\mean{N_{\T}}$.
This is corroborated by the fact that the expression~(\ref{eq:NCox}) for
$I_1(0)$
is convergent for $\th_1<\th_2$.
Using the estimate~(\ref{eq:th22}) for $1-\hat f(s)$ in (\ref{eq:nc1}), we
obtain
\beq\label{eq:prod}
\mean{\nt}\approx
\mean{N_{\T}}\,\frac{\sin\pi\th_2}{\pi\th_2}\,\left(\frac{t}{T_0}\right)^{\th_2}.
\eeq
As in region~A,
$\mean{\nt}$ has the form~(\ref{eq:Nnut}), i.e., it is asymptotically the
product of the mean number $\mean{N_{\T}}$ of internal events between two resettings
by the mean number of resettings in $(0,t)$, given in the present case
by~(\ref{eq:Ntrenew}).

This product structure extends to the entire asymptotic distribution of $\nt$.
This can be shown by estimating~(\ref{eq:key}) as follows.
Anticipating that typical values of~$\nt$ will be large,
we set $z=\e^{-p}$, and analyse the regime where $p$ is small.
To leading order as $p\to0$, the numerator $\hat\psi(z,s)$ can be replaced by
\be
\hat\psi(1,s)=\hat\Phi(s)=\frac{1-\hat f(s)}{s}\approx a_2s^{\th_2-1}.
\ee
The analysis of the denominator of~(\ref{eq:key}) requires some care.
We have
\bea
1-\hat\phi(z,s)
&=&\int_0^\infty\dd T\,f(T)\bigl(1-\e^{-sT}Z(z,T)\bigr)
\nonumber\\
&\approx&\int_0^\infty\dd T\,f(T)\bigl(1-\e^{-sT}+p\mean{N_T}\bigr)
\nonumber\\
&\approx& 1-\hat f(s)+p\mean{N_{\T}}
\nonumber\\
&\approx& a_2s^{\th_2}+p\mean{N_{\T}},
\eea
where we used the expansion $Z(z,T)=1-p\mean{N_T}+\cdots$.
Finally,~(\ref{eq:key}) reduces to
\be
\lap{t}\mean{\e^{-p\nt}}\approx\frac{a_2s^{\th_2-1}}{a_2s^{\th_2}+p\mean{N_{\T}}},
\ee
yielding for the distribution of $\nt$ in the continuum limit
\be
\lap{t}f_{\nt}(n)\approx\frac{a_2s^{\th_2-1}}{\mean{N_{\T}}}
\exp\left(-\frac{na_2s^{\th_2}}{\mean{N_{\T}}}\right).
\ee
Comparing this expression to~(\ref{eq:plap}),
we obtain the scaling result
\beq\label{eq:NcalC}
\nt\approx\mean{N_{\T}}\,\frac{\Xtwo}{\Gamma(1-\th_2)}\left(\frac{t}{T_0}\right)^{\th_2}
\approx\mean{N_{\T}}\,M_t,
\eeq
where the rescaled random variable $\Xtwo$ has density $f_{\Xtwo}(x)$,
given by~(\ref{eq:fX}).
This expression, which generalises (\ref{eq:prod}), shows that $\nt$ is,
asymptotically, equal to the product of the mean number $\mean{N_{\T}}$
of internal events between two resettings by the random number of resettings
(see~(\ref{eq:Ntscal})).

\subsubsection*{Region~D $(\th_2<\th_1<1)$}

In this region, the statistics of $\nt$ is more complex.
First, since both exponents are less than unity,
large fluctuations in $\nt$ are to be expected.
Second,
the number of internal events between any two consecutive resettings
is itself expected to diverge for late times.
The integrals $I_1(s)$ and $I_2(s)$ are indeed divergent as $s\to0$.
Their leading behaviour can be estimated by using
the expression~(\ref{eq:Ntrenew}) of $\mean{N_T}$,
where $\th$ is replaced by~$\th_1$.
We thus obtain
\be
I_1(s)+sI_2(s)\approx\frac{\sin\pi\th_1}{\pi\th_1}\,\frac{T_0^{\th_2}}{\tau_0^{\th_1}}\,
\Gamma(\th_1-\th_2)s^{\th_2-\th_1}.
\ee
Substituting this expression and the estimate~(\ref{eq:th22}) for $1-\hat f(s)$
into~(\ref{eq:nc1}), we obtain
\beq\label{eq:resN}
\mean{\nt}\approx
E_{\th_1,\th_2}
\,\frac{\sin\pi\th_1}{\pi\th_1}\,\left(\frac{t}{\tau_0}\right)^{\th_1},
\qquad
E_{\th_1,\th_2}=\frac{\Gamma(\th_1-\th_2)}{\Gamma(\th_1)\Gamma(1-\th_2)},
\eeq
which is the product of the enhancement factor $E_{\th_1,\th_2}$
by the mean number of events of the internal process
in the absence of resetting (see~(\ref{eq:Ntrenew})).
The enhancement factor becomes unity as $\th_2\to0$,
in which case (\ref{eq:resN}) gives back (\ref{eq:Ntrenew}),
where $\th$ is replaced by~$\th_1$.
This factor is an increasing function of $\th_2$, which diverges as $\th_2\to\th_1$
(see section~\ref{sec:regionDlim} for further discussion on these two limits).

The presence of the enhancement factor $E_{\th_1,\th_2}$,
which depends continuously on the two exponents~$\th_1$ and $\th_2$,
confirms that region~D is where the distribution of~$\nt$
exhibits the highest level of complexity.
The analysis of the entire asymptotic distribution of~$\nt$ in this region
is the focus of section~\ref{sec:regionD}.

We conclude this section by examining the behaviour of $\nt$
along the boundaries between the various regions of the phase diagram
shown in figure~\ref{fig:phase}, in order of increasing complexity.

\subsubsection*{Between regions~A and~B $(\th_2=1$ and $\th_1>1)$}

If $\th_2$ goes to unity from region~A,
both $\mean{N_{\T}}$ and $\mean{\T}$ diverge at the same pace,
and their ratio goes to the finite limit $1/\mean{\ttau}$,
hence the expressions~(\ref{eq:NregionA}) and~(\ref{eq:region3}) match smoothly.

\subsubsection*{Between regions~A and~C $(\th_2=1$ and $0<\th_1<1)$}

Along this phase boundary, $\mean{N_{\T}}$ is convergent,
and $\hat f(s)\approx1+T_0s\ln(T_\star s)$,
in analogy with~(\ref{eq:thone}).
Inserting these estimates into~(\ref{eq:nc1}),
we obtain
\be
\lap{t}\mean{\nt}\approx-\frac{\mean{N_{\T}}}{T_0s^2\ln(T_\star s)},
\ee
hence
\be
\mean{\nt}\approx\frac{\mean{N_{\T}}}{\ln(t/T_\star)+\gamma-1}\,\frac{t}{T_0}.
\ee
This can be rewritten as (see~(\ref{eq:singlemean}))
\be
\mean{\nt}\approx\frac{\mean{N_{\T}}}{\mean{\T_t}}\,t,
\ee
which matches with~(\ref{eq:NregionA}).
One can verify that typical fluctuations of $\nt$ around its mean value are
marginally negligible.
This property, which holds all over region~A,
is emerging in region~C as well in the limit $\th_2\to1$ (see~(\ref{eq:1})).

\subsubsection*{Between regions~B and~D $(\th_1=1$ and $0<\th_2<1)$}

Along this phase boundary,
the integrals $I_1(s)$ and $I_2(s)$ are divergent as $s\to0$.
Their leading behaviour can be estimated by inserting
the expression~(\ref{eq:nmargin}) of $\mean{N_T}$ into~(\ref{eq:II}).
We thus obtain
\be
I_1(s)+sI_2(s)\approx
-\frac{T_0^{\th_2}\Gamma(1-\th_2)}{\tau_0\,s^{1-\th_2}\ln(\tau_\star s)}.
\ee
Substituting this expression and the estimate~(\ref{eq:th22}) for $1-\hat f(s)$
into~(\ref{eq:nc1}), we obtain
\be
\lap{t}\mean{\nt}\approx-\frac{1}{\tau_0s^2\ln(\tau_\star s)},
\ee
hence
\be
\mean{\nt}\approx\frac{1}{\ln(t/\tau_\star)+\gamma-1}\,\frac{t}{\tau_0},
\ee
thus (see~(\ref{eq:singlemean}))
\be
\mean{\nt}\approx\frac{t}{\mean{\ttau_t}},
\ee
which matches with~(\ref{eq:region3}).

Typical fluctuations of $\nt$ around its mean are again marginally negligible.
This property, which holds all over region~B,
is emerging in region~D as well in the limit $\th_1\to1$ (see~(\ref{eq:th1one})).

\subsubsection*{Between regions~C and~D $(0<\th_1=\th_2<1)$}

Along this phase boundary, the integral~$I_1(s)$ is logarithmically divergent,
whereas $I_2(s)$ can be neglected.
Denoting by $\th$ the common value of $\th_1$ and $\th_2$,
and inserting the expressions~(\ref{eq:Ntrenew}) and~(\ref{eq:t1t2})
into~(\ref{eq:II}),
we obtain
\be
\lap{t}\mean{\nt}\approx-\frac{\sin\pi\th}{\pi\Gamma(1-\th)\tau_0^\th}\,
\frac{\ln s}{s^{1+\th}},
\ee
where the finite part of the logarithm has no simple expression in general
and will therefore be omitted.
This yields
\beq\label{eq:nxave}
\mean{\nt}\approx\frac{\sin^2\pi\th}{\pi^2\th}\left(\frac{t}{\tau_0}\right)^\th\,\ln t.
\eeq
At variance with all other phase boundaries,
the number $\nt$ of internal events keeps fluctuating in the present situation.
It is indeed clear that $\nt$ is proportional to the reduced random variable $X_\th$
as the phase boundary is approached from either side.
This property, which holds all over region~C,
is emerging in region~D as well in the limit $\th_2\to\th_1$ (see~(\ref{eq:ytox})).
The proportionality constant between $\nt$ and $X_\th$ is determined by comparing
the expressions~(\ref{eq:nxave}) of $\mean{\nt}$
and~(\ref{eq:momentX}) of $\mean{X_\th}$.
We thus obtain the asymptotic estimate
\be
\nt\approx\frac{\sin\pi\th}{\pi}\frac{X_\th}{\Gamma(1-\th)}
\left(\frac{t}{\tau_0}\right)^\th\,\ln t
\ee
along the boundary between regions~C and~D.

\subsubsection*{The quadruple point $(\th_1=\th_2=1)$}

Skipping every detail, we mention that the mean number of internal events
scales as
\be
\mean{\nt}\approx\frac{\ln\ln t}{\ln t}\,\frac{t}{\tau_0}
\ee
at the quadruple point where the four regions of the phase diagram meet.

\section{Asymptotic distribution of $\nt$ in region~D ($0<\th_2<\th_1<1$)}
\label{sec:regionD}

The growth law~(\ref{eq:resN}) of the mean number $\mean{\nt}$
of internal events in region~D suggests to postulate the scaling form
\beq\label{eq:y4sca}
\nt\approx\frac{\Y}{\Gamma(1-\th_1)}\left(\frac{t}{\tau_0}\right)^{\th_1},
\eeq
in analogy with~(\ref{eq:Ntscal}),
where the rescaled random variable $\Y$ has a non-trivial universal distribution
depending only on the two exponents $\th_1$ and $\th_2$,
whose density will be denoted by $f_{\Y}(y)$.
The current section is dedicated to a thorough examination of this probability density.

\subsection{Fundamental integral equation}

Our starting point is again the exact expression~(\ref{eq:key}).
As in section~\ref{sec:diagram}, anticipating that typical values of~$\nt$ are
large for late times,
we set $z=\e^{-p}$,
and analyse the regime where $p$ is small.
The scaling form~(\ref{eq:Ntscal}) of $N_t$ implies that the generating
function $Z(z,t)$ defined in~(\ref{eq:ZNt}) scales as
\beq\label{eq:Zth1}
Z(z,t)\approx\hat f_{\Xone}(u),\qquad u=\frac{p}{\Gamma(1-\th_1)}
\left(\frac{t}{\tau_0}\right)^{\th_1}
\eeq
(see~(\ref{eq:mittag})) in the regime where $p$ is small and $t$ is large.

Inserting the tail expressions~(\ref{eq:th21}) and the scaling
form~(\ref{eq:Zth1}) into the expression~(\ref{eq:psidef})
for $\hat\psi(z,s)$,
and changing the integration variable from~$T$ to the corresponding rescaled
variable~$u$,
we obtain the scaling form
\beq\label{eq:psiasy}
\hat\psi(z,s)\approx T_0^{\th_2}s^{\th_2-1}\,I(\lam),
\eeq
where
\beq\label{eq:lamdef}
\lam=\tau_0s\left(\frac{\Gamma(1-\th_1)}{p}\right)^{1/\th_1}
\eeq
and
\beq\label{eq:idef}
I(\lam)=\frac{\lam^{1-\th_2}}{\th_1}\int_0^\infty\dd u
\,\e^{-\lam u^{1/\th_1}}u^{(1-\th_2)/\th_1-1}\hat f_{\Xone}(u).
\eeq
Using the definition~(\ref{eq:phidef}) of $\hat\phi(z,s)$, the denominator
of~(\ref{eq:key}) can be written as
\bea
1-\hat\phi(z,s)
&=&1-\int_0^\infty\dd T\,f(T)\,\e^{-sT}Z(z,T)
\nonumber\\
&=&1-\hat f(s)+\int_0^\infty\dd T\,f(T)\,\e^{-sT}\bigl(1-Z(z,T)\bigr).
\eea
Using again~(\ref{eq:th21}) and~(\ref{eq:Zth1}), we obtain the scaling form
\beq\label{eq:phiasy}
1-\hat\phi(z,s)\approx(T_0s)^{\th_2}\,K(\lam),
\eeq
with
\beq\label{eq:kdef}
K(\lam)=\Gamma(1-\th_2)+\th_2\,J(\lam),
\eeq
and
\beq\label{eq:jdef}
J(\lam)=\frac{\lam^{-\th_2}}{\th_1}\int_0^\infty\dd u
\,\e^{-\lam u^{1/\th_1}}u^{-1-\th_2/\th_1}\bigl(1-\hat f_{\Xone}(u)\bigr).
\eeq
Finally, inserting~(\ref{eq:psiasy}) and~(\ref{eq:phiasy}) into~(\ref{eq:key}),
we are left with the estimate
\beq\label{eq:zes1}
\hat\Z(z,s)\approx\frac{I(\lam)}{s\,K(\lam)}.
\eeq
On the other hand,
the postulated scaling law~(\ref{eq:y4sca})
yields
\beq\label{eq:hatZcal2}
\hat\Z(z,s)\approx\int_0^\infty\dd t\,\e^{-st}
\int_0^\infty\dd y\,\e^{-uy}f_{\Y}(y),
\eeq
where $u$ is defined in~(\ref{eq:Zth1}).
Using the rescaled variable $\lam$ introduced in~(\ref{eq:lamdef}),
as well as $\mu=u^{1/\th_1}$, whereby $st=\lam\mu$,
we have
\beq\label{eq:zes2}
\hat\Z(z,s)\approx\frac{\lam}{s}\int_0^\infty\dd\mu\,\e^{-\lam\mu}
\int_0^\infty\dd y\,\e^{-\mu^{\th_1}y}f_{\Y}(y).
\eeq
A comparison between~(\ref{eq:zes1}) and~(\ref{eq:zes2})
corroborates the scaling form~(\ref{eq:y4sca})
and yields an integral equation for the density $f_{\Y}(y)$, of the form
\beq\label{eq:fundam}
\L(\lam)=\R(\lam),
\eeq
with, on the left-hand side,
\beq\label{eq:ldef}
\L(\lam)=\lam\int_0^\infty\dd\mu\,\e^{-\lam\mu}\int_0^\infty\dd y\,\e^{-\mu^{\th_1}y} f_{\Y}(y),
\eeq
and, on the right-hand side,
\beq\label{eq:rdef}
\R(\lam)=\frac{I(\lam)}{K(\lam)}.
\eeq

The fundamental equation~(\ref{eq:fundam})
is the starting point of the analysis that follows,
where we successively investigate the moments of $\Y$
(section~\ref{sec:region4moms}),
the behaviour of its probability density $f_{\Y}(y)$ at small and large~$y$
(sections~\ref{sec:region4small} and~\ref{sec:region4large}),
the three limiting situations corresponding to the edges of region~D
(section~\ref{sec:regionDlim}),
a measure of the fluctuations (section~\ref{sec:region4var}),
and finally an integral representation of $f_{\Y}(y)$
(section~\ref{sec:region4rep}).

\subsection{Moments of $\,\Y$}
\label{sec:region4moms}

The moments of $\Y$ can be extracted from~(\ref{eq:fundam}) as follows.
Consider first its left-hand side $\L(\lam)$ given by~(\ref{eq:ldef}).
Expanding the exponential in the innermost integral as a power series
in~$\mu^{\th_1}$ and performing the integrals,
we can recast $\L(\lam)$ as a power series
in the variable
\beq\label{eq:zetadef}
\z=-\lam^{-\th_1},
\eeq
reading
\beq\label{eq:lser}
\L(\lam)\equiv\tilde\L(\z)=\sum_{k\ge0}\frac{\Gamma(k\th_1+1)}{k!}\,\mean{\Y^k}\,\z^k.
\eeq
Similarly, by inserting the power series~(\ref{eq:mittag}) for $\hat
f_{\Xone}(u)$
into~(\ref{eq:idef}) and~(\ref{eq:jdef}) and performing the integrals,
we obtain the following power series in $\z$:
\beqa
I(\lam)
&\equiv&\tilde I(\z)
=\sum_{k\ge0}\frac{\Gamma(k\th_1+1-\th_2)}{\Gamma(k\th_1+1)}\,\z^k,
\label{eq:iser}
\\
K(\lam)
&\equiv&\tilde K(\z)
=-\th_2\sum_{k\ge0}\frac{\Gamma(k\th_1-\th_2)}{\Gamma(k\th_1+1)}\,\z^k.
\label{eq:kser}
\eeqa

The functions $\tilde I(\z)$ and $\tilde K(\z)$ are simple generalisations of the Wright function (see~\cite{W} and references therein).
They are analytic in the complex $\z$-plane cut along the positive real axis from 1 to $\infty$.
They are related by the differential identity
\be
\tilde I(\z)=\tilde K(\z)-\frac{\th_1}{\th_2}\,\z\tilde K'(\z),
\ee
where the accent denotes a differentiation, entailing that (see~(\ref{eq:rdef}))
\beq\label{eq:Rzeta}
\R(\lam)\equiv\tilde\R(\z)=\frac{\tilde I(\z)}{\tilde K(\z)}
=1-\frac{\th_1}{\th_2}\,\frac{\z\tilde K'(\z)}{\tilde K(\z)}
\eeq
is nearly a logarithmic derivative.

By identifying the coefficients of successive powers of $\z$ in the power
series
$\tilde\L(\z)$ (see~(\ref{eq:lser})) and $\tilde\R(\z)$
(see~(\ref{eq:iser}),~(\ref{eq:kser}),~(\ref{eq:Rzeta})),
we obtain explicit expressions for the moments of $\Y$,
depending only on the exponents $\th_1$ and $\th_2$:
\bea
\mean{\Y}
&=&\frac{\th_1\,\Gamma(\th_1-\th_2)}{\Gamma(\th_1+1)^2\Gamma(1-\th_2)},
\nonumber\\
\mean{\Y^2}
&=&\frac{4\th_1\,\Gamma(2\th_1-\th_2)}{\Gamma(2\th_1+1)^2\Gamma(1-\th_2)}
\nonumber\\
&+&\frac{2\th_1\th_2\,\Gamma(\th_1-\th_2)^2}
{\Gamma(\th_1+1)^2\Gamma(2\th_1+1)\Gamma(1-\th_2)^2},
\nonumber\\
\mean{\Y^3}
&=&\frac{18\th_1\,\Gamma(3\th_1-\th_2)}{\Gamma(3\th_1+1)^2\Gamma(1-\th_2)}
\nonumber\\
&+&\frac{18\th_1\th_2\,\Gamma(\th_1-\th_2)\Gamma(2\th_1-\th_2)}
{\Gamma(\th_1+1)\Gamma(2\th_1+1)\Gamma(3\th_1+1)\Gamma(1-\th_2)^2}
\nonumber\\
&+&\frac{6\th_1\th_2^2\,\Gamma(\th_1-\th_2)^3}
{\Gamma(\th_1+1)^3\Gamma(3\th_1+1)\Gamma(1-\th_2)^3},
\eea
and so on.
The expression for $\mean{\Y}$ given above is in accordance
with~(\ref{eq:resN}).
Higher moments have expressions of increasing complexity,
involving gamma functions of more and more different arguments.

In the regime where $\th_1$ and $\th_2$ simultaneously go to 0,
the moments of $\Y$ maintain a non-trivial rational dependence on the ratio
$\aa=\th_2/\th_1$, such that $0<\aa<1$.
The preceding expressions indeed reduce to
\beqa\label{eq:moments}
\lim_{\th_1,\th_2\to0}\mean{\Y}
&=&\frac{1}{1-\aa},
\nonumber\\
\lim_{\th_1,\th_2\to0}\mean{\Y^2}
&=&\frac{2(2-2\aa+\aa^2)}{(2-\aa)(1-\aa)^2},
\nonumber\\
\lim_{\th_1,\th_2\to0}\mean{\Y^3}
&=&\frac{6(6-12\aa+12\aa^2-5\aa^3+\aa^4)}{(3-\aa)(2-\aa)(1-\aa)^3}.
\eeqa

\subsection{Behaviour at small values of $\Y$}
\label{sec:region4small}

The behaviour of the density $f_{\Y}(y)$ at small $y$ can be extracted from~(\ref{eq:fundam})
by noticing that $y\to0$ corresponds to $\mu\to\infty$ and to $\lam\to0$ (see~(\ref{eq:ldef})).

Let us assume provisionally that the exponents $\th_1$ and $\th_2$
obey the inequality $\th_1+\th_2<1$.
The behaviour of $I(\lam)$ at small $\lam$ can be obtained by inserting
into~(\ref{eq:idef})
the behaviour of $\hat f_{\Xone}(u)$ at large $u$, namely
(see~(\ref{eq:fsmall}))
\be
\hat f_{\Xone}(u)\approx\frac{1}{\Gamma(1-\th_1)u}.
\ee
We thus obtain
\be
I(\lam)\approx\frac{\Gamma(1-\th_1-\th_2)}{\Gamma(1-\th_1)}\,\lam^{\th_1}.
\ee
The behaviour of $J(\lam)$ and $K(\lam)$ at small $\lam$ read
(see~(\ref{eq:kdef}),~(\ref{eq:jdef}))
\be
J(\lam)\approx C\lam^{-\th_2},\qquad
K(\lam)\approx C\th_2\lam^{-\th_2},
\ee
with
\be
C=\frac{1}{\th_1}\int_0^\infty\dd u
\,\e^{-\lam u^{1/\th_1}}u^{-1-\th_2/\th_1}\bigl(1-\hat f_{\Xone}(u)\bigr).
\ee
An integration by parts followed by some algebra yields
\be
C=\frac{\Gamma(1-\th_2/\th_1)}{\th_2}\,\mean{\Xone^{\th_2/\th_1}},
\ee
where $\Xone$ has density $f_{\Xone}(x)$ (see~(\ref{eq:fX})).
Extending the moment formula~(\ref{eq:momentX}) to the non-integer value $k=\th_2/\th_1$,
we obtain
\be
C=\frac{\Gamma(1-\th_2/\th_1)\Gamma(1+\th_2/\th_1)}{\th_2\Gamma(1+\th_2)},
\ee
and finally
\beq\label{eq:rasy}
\R(\lam)\approx\frac{\Gamma(1-\th_1-\th_2)\Gamma(1+\th_2)}
{\Gamma(1-\th_1)\Gamma(1-\th_2/\th_1)\Gamma(1+\th_2/\th_1)}\,\lam^{\th_1+\th_2},
\eeq
as long as the inequality $\th_1+\th_2<1$ holds.
When this inequality is not obeyed,
the above singular term is still there,
but it is subleading with respect to a regular term linear in $\lam$.

The power-law singularity in~(\ref{eq:rasy}) suggests to assume the power-law behaviour
\beq\label{eq:fysmall}
f_{\Y}(y)\approx Ay^\omega\qquad(y\to0).
\eeq
Inserting this scaling expression into~(\ref{eq:ldef}), performing the integrals,
and identifying exponents and amplitudes with~(\ref{eq:rasy}) yields
\be
\omega=\frac{\th_2}{\th_1}
\ee
and
\be
A=\frac{\Gamma(1+\th_2)}{\Gamma(1-\th_1)\Gamma(1-\th_2/\th_1)\Gamma(1+\th_2/\th_1)^2}.
\ee
The power-law behaviour~(\ref{eq:fysmall})
of the density of $\Y$ as $y\to0$ is different from that of the density of $\X$,
which goes to a finite constant as $x\to0$ (see~(\ref{eq:fsmall})).
However, when $\th_2\to0$, the exponent $\omega$ tends to zero,
and the amplitude $A$ matches with this constant.

\subsection{Behaviour at large values of $\Y$}
\label{sec:region4large}

In order to analyse the tail behaviour of the density $f_{\Y}(y)$ for large values of $y$,
we use the property that the latter behaviour is related to the asymptotic growth law
of the moments $\mean{\Y^k}$ at large $k$.
The latter can be estimated by considering
the denominator $\tilde K(\z)$ of the right-hand side $\tilde\R(\z)$ of~(\ref{eq:fundam}),
given by the power series~(\ref{eq:kser}).
As $\z$ increases from 0 to 1, i.e., its radius of convergence,
$\tilde K(\z)$ decreases from the positive value $\tilde K(0)=\Gamma(1-\th_2)$
to some finite negative value~$\tilde K(1)$.
There is therefore a critical value of $\z$, denoted by $\z_c$, such that
\beq\label{eq:zcdef}
\tilde K(\z_c)=0.
\eeq
Both $\tilde R(\z)$ and $\tilde L(\z)$ therefore have a simple pole at $\z_c$.
As a consequence of~(\ref{eq:lser}), we have
\beq\label{eq:ymoms}
\mean{\Y^k}\sim\frac{k!}{\Gamma(k\th_1+1)\,\z_c^k},
\eeq
up to an inessential constant.
Comparing the asymptotic estimate~(\ref{eq:ymoms}) to the exact
expression~(\ref{eq:momentX})
of the moments of $\X$,
we are led to the conclusion that the tail behaviour of $f_{\Y}(y)$
is obtained by replacing $\th$ by $\th_1$ and $x$ by the product $\z_c y$
in the compressed exponential estimate~(\ref{eq:flarge}),
obtaining thus
\beq\label{eq:ylarge}
f_{\Y}(y)\sim\exp\left(-(1-\th_1)(\th_1^{\th_1}\z_c y)^{1/(1-\th_1)}\right).
\eeq

\begin{figure}
\begin{center}
\includegraphics[angle=0,width=.7\linewidth,clip=true]{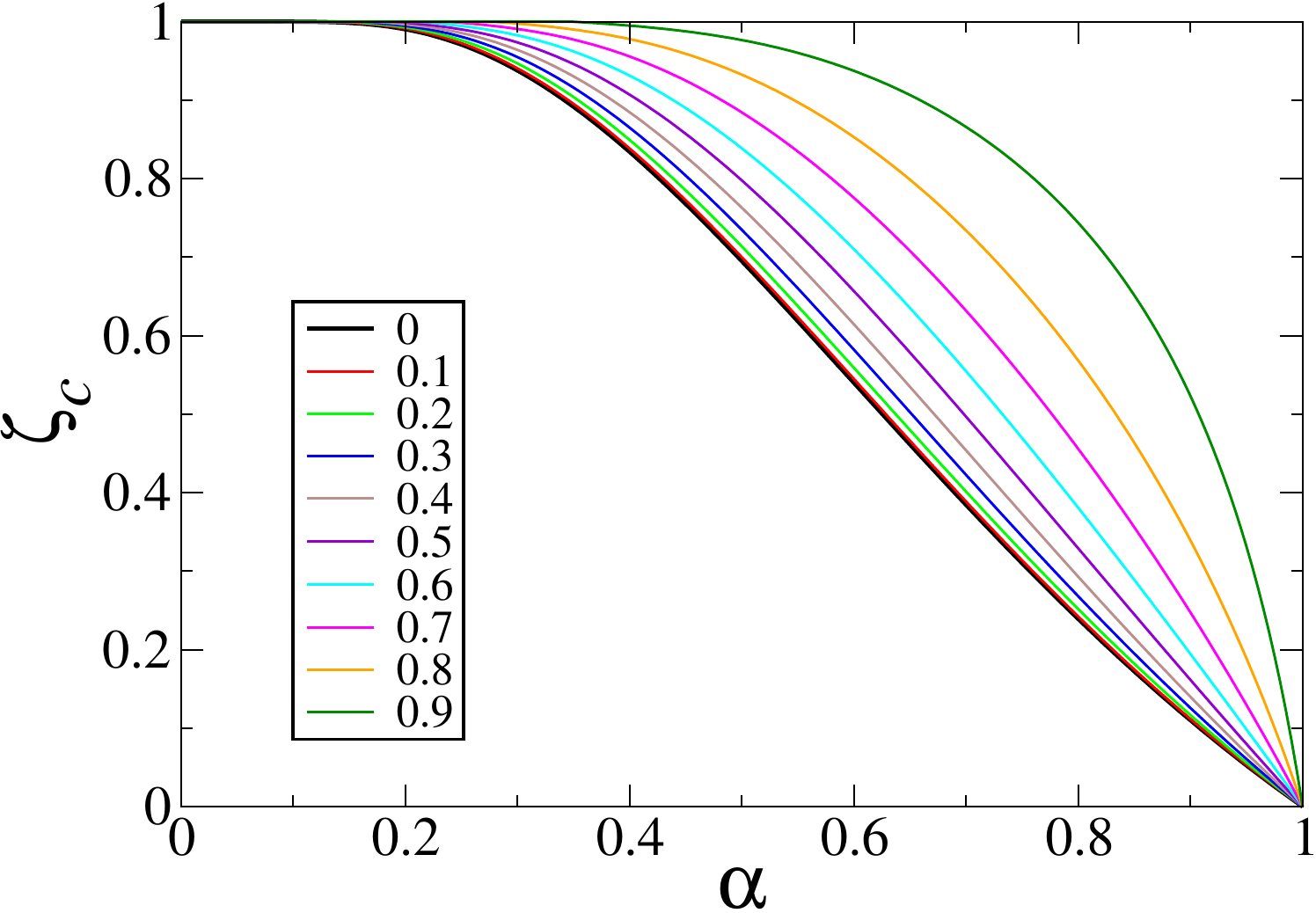}
\caption{\small
Quantity $\z_c$ defined by~(\ref{eq:zcdef})
and entering the estimates~(\ref{eq:ymoms}) and~(\ref{eq:ylarge}),
plotted against the ratio $\aa=\th_2/\th_1$ for several values of $\th_1$
(see legend).}
\label{fig:zetaplot}
\end{center}
\end{figure}

This result depends on $\th_2$ only through the quantity $\z_c$,
defined in~(\ref{eq:zcdef}).
The latter has a non-trivial dependence on $\th_1$ and $\th_2$,
decreasing from 1 to 0 as~$\th_2$ is increased from 0 to $\th_1$.
It is plotted in figure~\ref{fig:zetaplot}
against the ratio $\aa=\th_2/\th_1$ for several values of $\th_1$ (see legend).
For $\th_2\to0$, it can be argued that $\z_c$ departs from unity
with an exponentially small singularity of the form
\beq\label{eq:zcsmall}
1-\z_c\sim\exp\left(-\frac{\abs{\ln(1-\th_1)}}{\th_2}\right).
\eeq
For $\th_2\to\th_1$, the estimate~(\ref{eq:keps}), to be derived below,
yields the linear behaviour
\beq\label{eq:zclarge}
\z_c\approx\frac{\pi}{\sin\pi\th_1}\,(\th_1-\th_2).
\eeq
In the regime where $\th_1$ and $\th_2$ simultaneously go to 0,
the series~(\ref{eq:kser}) reduces~to
\be
\tilde K(\z)=-\aa\sum_{k\ge0}\frac{\z^k}{k-\aa},
\ee
with $\aa=\th_2/\th_1$,
so that $\z_c$ keeps a non-trivial dependence on the ratio $\aa$
(thick black curve in figure~\ref{fig:zetaplot}).
The estimates~(\ref{eq:zcsmall}) and~(\ref{eq:zclarge})
can be made more precise in this regime:
\bea
1-\z_c&=&\exp\left(-\frac{1}{\aa}+\frac{\pi^2\aa}{6}+\cdots\right)\qquad(\aa\to0),
\\
\z_c&=&(1-\aa)+(1-\aa)^2+\cdots\qquad(\aa\to1).
\eea

\subsection{Three limiting situations}
\label{sec:regionDlim}

The three limiting situations of interest correspond to the edges of the
triangular region~D (see figure~\ref{fig:phase}), that is, $\th_2\to0$,
$\th_1\to1$, and $\th_2\to\th_1$.

\subsubsection*{Limit $\th_2\to0$}

In this limit, the series~(\ref{eq:iser}) and~(\ref{eq:kser}) reduce to
\beq\label{eq:rzero}
\tilde I(\z)=\frac{1}{1-\z},\qquad
\tilde K(\z)=1,\qquad\tilde\R(\z)=\frac{1}{1-\z},
\eeq
thus~(\ref{eq:lser}) yields
\be
\mean{Y_{\th_1,0}^k}=\frac{k!}{\Gamma(k\th_1+1)}.
\ee
A comparison with~(\ref{eq:momentX}) leads to the conclusion that, in this limit,
the distribution of $\Y$ becomes that of $\Xone$ (see~(\ref{eq:fX})).
In other words,
\beq\label{eq:fzero}
\lim_{\th_2\to0}f_{\Y}(y)=f_{\Xone}(y).
\eeq
This result can be understood as follows.
When $\th_2\to0$, the distribution of the time intervals $\T_1,\T_2,\dots$ becomes very broad,
entailing that the largest of them, $\T_\max=\max(\T_1,\T_2,\dots)$,
nearly spans the whole time interval $(0,t)$.
More precisely, using the expression given in~\cite[eq.~(3.34)]{gms} for the limiting ratio
\be
\lim_{t\to\infty}\frac{1}{t}\mean{\T_\max}
=\int_0^\infty\frac{\dd x}{1+x^\th\e^x\int_0^x{\rm d}u\,u^{-\th}\e^{-u}},
\ee
it can be checked that this ratio goes to unity
as $\th\to0$ (where $\th$ stands for $\th_2$ in the present context),
meaning that the entire process simplifies to the internal process
over a time interval $\T_\max\approx t$.
See also the comment below~(\ref{eq:resN}) on this limit.

\subsubsection*{Limit $\th_1\to1$}

In this limit,
by inserting $\hat f_1(u)=\e^{-u}$ (see~(\ref{eq:1}))
into the integrals~(\ref{eq:idef}) and~(\ref{eq:jdef}),
and performing the latter integrals,
we obtain, after some algebra,
\be
\R(\lam)=\frac{\lam}{\lam+1},\qquad
\tilde\R(\z)=\frac{1}{1-\z},
\ee
whereby~(\ref{eq:lser}) yields
\be
\mean{Y_{1,\th_2}^k}=1.
\ee
In the limit $\th_1\to1$,
the distribution of $\Y$ thus becomes degenerate:
\beq\label{eq:th1one}
\lim_{\th_1\to1}f_{\Y}(y)=\delta(y-1),
\eeq
regardless of the value of $\th_2$.
The scaling variable $\X$ has the same degenerate distribution as $\th\to1$
(see~(\ref{eq:1})).

\subsubsection*{Limit $\th_2\to\th_1$}

In this limit, the series~(\ref{eq:kser}) become singular,
in the sense that the term corresponding to $k=1$ diverges.
Setting
\be
\th_2=\th_1-\eps,
\ee
we obtain
\beqa
\tilde I(\z)&\approx&\Gamma(1-\th_1),
\nonumber\\
\tilde K(\z)&\approx&\Gamma(1-\th_1)-\frac{\z}{\Gamma(\th_1)\,\eps},
\label{eq:keps}
\\
\tilde R(\z)&\approx&\frad{1}{1-\frad{\sin\pi\th_1}{\pi\eps}\,\z},
\nonumber
\eeqa
whereby~(\ref{eq:lser}) yields
\be
\mean{Y_{\th_1,\th_1-\eps}^k}\approx\left(\frac{\sin\pi\th_1}{\pi\eps}\right)^k\,
\frac{k!}{\Gamma(k\th_1+1)}.
\ee
A comparison with~(\ref{eq:momentX}) leads to the following equivalence
\beq\label{eq:ytox}
Y_{\th_1,\th_1-\eps}\approx\frac{\sin\pi\th_1}{\pi\eps}\,\Xone,
\eeq
to leading order as $\eps=\th_1-\th_2\to0$,
where $\Xone$ is distributed according to~(\ref{eq:fX}), with exponent $\th_1$.
In other words,
\be
f_{Y_{\th_1,\theta_1-\eps}}(y)\approx\frac{\pi\eps}{\sin\pi\th_1}\;
f_{\Xone}\!\!\left(\frac{\pi\eps}{\sin\pi\th_1}\,y\right).
\ee

\subsection{A measure of fluctuations}
\label{sec:region4var}

In order to obtain a quantitative measure of the size of the fluctuations of $\Y$, we consider its reduced variance, denoted by
\beq\label{eq:vdef}
V=\frac{\var\Y}{\mean{\Y}^2}=\frac{\mean{\Y^2}}{\mean{\Y}^2}-1.
\eeq
The explicit expressions~(\ref{eq:ymoms}) of the first two moments of $Y$ yield
\beq\label{eq:v}
V=\frac{4\Gamma(\th_1+1)^4\Gamma(1-\th_2)\Gamma(2\th_1-\th_2)}
{\th_1\Gamma(2\th_1+1)^2\Gamma(\th_1-\th_2)^2}
+\frac{2\th_2\Gamma(\th_1+1)^2}{\th_1\Gamma(2\th_1+1)}-1.
\eeq

For a fixed $\th_1$, $V$ takes its maximal value,
\beq\label{eq:vmax}
V_\max=\frac{2\Gamma(\th_1+1)^2}{\Gamma(2\th_1+1)}-1
=\frac{\mean{X_{\th_1}^2}}{\mean{X_{\th_1}}^2}-1
\eeq
at both endpoints of region~D,
namely for $\th_2\to0$ and $\th_2\to\th_1$,
where $\Y$ becomes proportional to $X_{\th_1}$.

\begin{figure}
\begin{center}
\includegraphics[angle=0,width=.7\linewidth,clip=true]{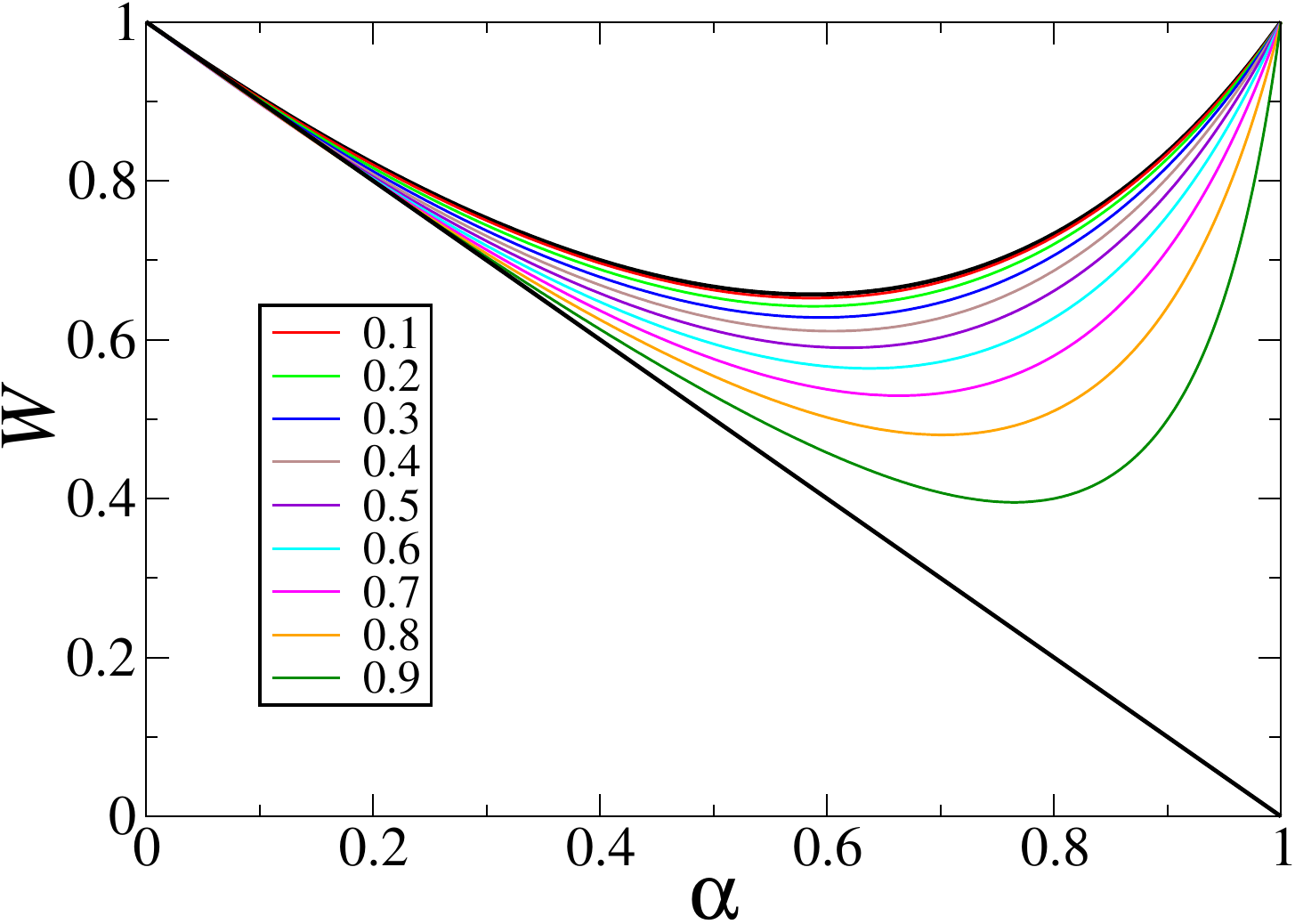}
\caption{\small
Ratio $W=V/V_\max$,
where $V$ is the reduced variance of the scaling variable $\Y$ (see~(\ref{eq:vdef})),
and $V_\max$ is given by~(\ref{eq:vmax}),
plotted against the ratio $\aa=\th_2/\th_1$ for several values of $\th_1$ (see legend).
Thick black curves: limit expressions~(\ref{eq:w0}) and~(\ref{eq:w1}).}
\label{fig:wplot}
\end{center}
\end{figure}

For intermediate values of $\th_2$, the distribution of $\Y$ has less
pronounced fluctuations,
testified by a smaller value of the reduced variance $V$.
Figure~\ref{fig:wplot} shows plots of the ratio
\be
W=\frac{V}{V_\max}
\ee
against the ratio $\aa=\th_2/\th_1$ for several values of $\th_1$ (see legend).
This quantity starts decreasing from its initial value $W=1$ for $\aa=0$,
goes through a minimum, and increases back to $W=1$ for $\aa=1$.
The minimum gets deeper and deeper as $\th_1$ is increased from 0 to 1.
In the regime where $\th_1$ and $\th_2$ simultaneously go to 0,
the expression~(\ref{eq:v}) reduces to
\beq\label{eq:w0}
W=\frac{2-3\aa+2\aa^2}{2-\aa}
=1-\frac{2\aa(1-\aa)}{2-\aa},
\eeq
in accord with~(\ref{eq:moments}).
This expression reaches a non-trivial minimum $W_\min=4\sqrt{2}-5=0.656854\dots$ for
$\aa=2-\sqrt{2}=0.585786\dots$
In the opposite limit ($\th_1\to1$), we have
\beq\label{eq:w1}
W=1-\aa,
\eeq
testifying that the limits $\th_1\to1$ and $\th_2\to\th_1$ do not commute.

\subsection{Integral representation of the density}
\label{sec:region4rep}

\begin{figure}
\begin{center}
\includegraphics[angle=0,width=.5\linewidth,clip=true]{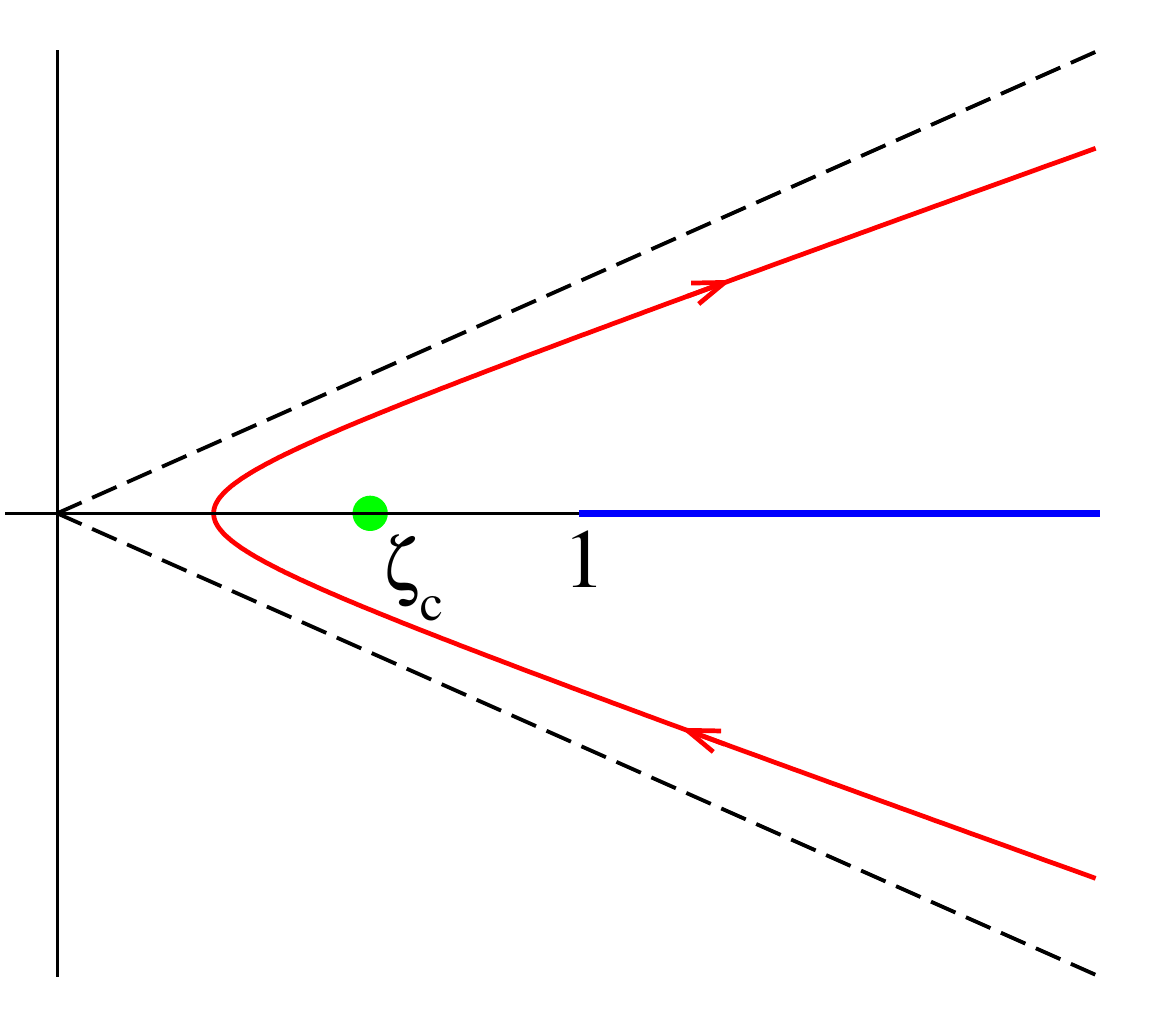}
\caption{\small
The complex $\z$-plane.
Green symbol: pole of $\tilde\R(\z)$ at $\z=\z_c$.
Blue line: branch cut of $\tilde\R(\z)$ from 1 to $\infty$.
Black dashed lines are at angles $\pm(1-\th_1)\pi/2$
beyond which the density $f_{\Xone}(\z y)$ looses its exponential decay.
Red curve: Integration contour $\Gamma$ to be used in the
representation~(\ref{eq:fint}).}
\label{fig:contour}
\end{center}
\end{figure}

The fundamental equation~(\ref{eq:fundam})
also yields an integral representation of the density $f_{\Y}(y)$.
Its left-hand side has the structure of two nested Laplace transforms.
Successively inverting these two transforms yields formally
\be
f_{\Y}(y)=\th_1\int\frac{\dd\lam}{2\pi\ii\lam}\,
\int\frac{\dd\mu}{2\pi\ii}\,\mu^{\th_1-1}\e^{\lam\mu+\mu^{\th_1}y}\R(\lam).
\ee
Changing integration variables from $\mu$ to $z=\lam\mu$
and from $\lam$ to~$\z$ (see~(\ref{eq:zetadef})),
we obtain
\be
f_{\Y}(y)=\int\frac{\dd\z}{2\pi\ii}\,
\underbrace{\int\frac{\dd z}{2\pi\ii}\,z^{\th_1-1}\e^{z-z^{\th_1}\z\, y}}\,\tilde\R(\z).
\ee
The expression underlined with a brace is merely the integral representation~(\ref{eq:fX})
of the density $f_{\Xone}(x)$, up to the replacement of the variable $x$ by the product $\z y$.
We have thus established the integral formula
\beq\label{eq:fint}
f_{\Y}(y)=\int_\Gamma\frac{\dd\z}{2\pi\ii}\,f_{\Xone}(\z y)\,\tilde\R(\z),
\eeq
which represents the density $f_{\Y}(y)$
as a continuous superposition of densities of the type $f_{\Xone}(x)$,
with weight $\tilde\R(\z)$ given in~(\ref{eq:Rzeta}).
The integration contour $\Gamma$ is described below and shown in figure~\ref{fig:contour}.

Let us start by considering the limit of~(\ref{eq:fint}) as $\th_2\to0$.
The expression~(\ref{eq:rzero}) shows that $\tilde\R(\z)$ has a simple pole at $\z=1$,
with residue $-1$.
In order to recover~(\ref{eq:fzero}),
the integration contour $\Gamma$ must encircle this pole once in the clockwise direction.

In the generic situation $(0<\th_2<\th_1<1)$,
$\tilde\R(\z)$ has two singularities on the positive real axis,
specifically a simple pole at $\z=\z_c$ between 0 and 1 and a branch cut
extending from 1 to $\infty$, as shown in figure~\ref{fig:contour}.
The density $f_{\Xone}(\z y)$ has the compressed exponential decay~(\ref{eq:flarge})
as long as the real part of $(\z y)^{1/(1-\th_1)}$ is positive,
i.e., $\z$ stays between the lines at angles $\pm(1-\th_1)\pi/2$.
The integration contour $\Gamma$ entering~(\ref{eq:fint}) should thus be placed
as shown in the figure.

The integral representation~(\ref{eq:fint}) is very appealing conceptually.
It is however not easy to handle in practice.
The specific techniques introduced earlier are indeed far more efficient
when exploring diverse facets of the distribution of $\Y$, such as
the calculation of moments, the behaviour of the density $f_{\Y}(y)$ at small
values of~$y$, and the three limiting situations associated with the edges of region~D.

The representation~(\ref{eq:fint}) simplifies in the special case where $\th_1=1/2$.
First of all, the density $f_{X_{1/2}}(x)$ has the simple form~(\ref{eq:12}).
The power series $\tilde I(\z)$ and~$\tilde K(\z)$
(see~(\ref{eq:iser}),~(\ref{eq:kser})) also become simpler.
The arguments of all gamma functions are of the form $k/2+c$ for various $c$.
This suggests to consider separately even ($k=2n$) and odd ($k=2n+1$) values of $k$.
We thus obtain
\beqa\label{eq:hypers}
\tilde I_\even(\z)
&=&\sum_{n\ge0}\frac{\Gamma(n+1-\th_2)}{n!}\,\z^{2n}
=\Gamma(1-\th_2)(1-\z^2)^{\th_2-1},
\\
\tilde I_\odd(\z)
&=&\sum_{n\ge0}\frac{\Gamma(n+\frac32-\th_2)}{\Gamma(n+\frac32)}\,\z^{2n+1}
=\frac{\Gamma(\frac32-\th_2)}{\Gamma(\frac32)}\,
\z F\!\left(\frac32-\th_2,1;\frac32;\z^2\right),
\nonumber\\
\tilde K_\even(\z)
&=&-\th_2\sum_{n\ge0}\frac{\Gamma(n-\th_2)}{n!}\,\z^{2n}
=\Gamma(1-\th_2)(1-\z^2)^{\th_2},
\nonumber\\
\tilde K_\odd(\z)
&=&-\th_2\sum_{n\ge0}\frac{\Gamma(n+\frac12-\th_2)}{\Gamma(n+\frac32)}\,\z^{2n+1}
=-\frac{\th_2\Gamma(\frac12-\th_2)}{\Gamma(\frac32)}\,
\z F\!\left(\frac12-\th_2,1;\frac32;\z^2\right),
\nonumber
\eeqa
where $F(a,b;c;z)$ is the hypergeometric series.
Thus~(\ref{eq:Rzeta}) reads
\beq\label{eq:Rzeta+}
\tilde\R(\z)=\frac{\tilde I_\even(\z)+\tilde I_\odd(\z)}{\tilde
K_\even(\z)+\tilde K_\odd(\z)}.
\eeq

\begin{figure}
\begin{center}
\includegraphics[angle=0,width=.75\linewidth,clip=true]{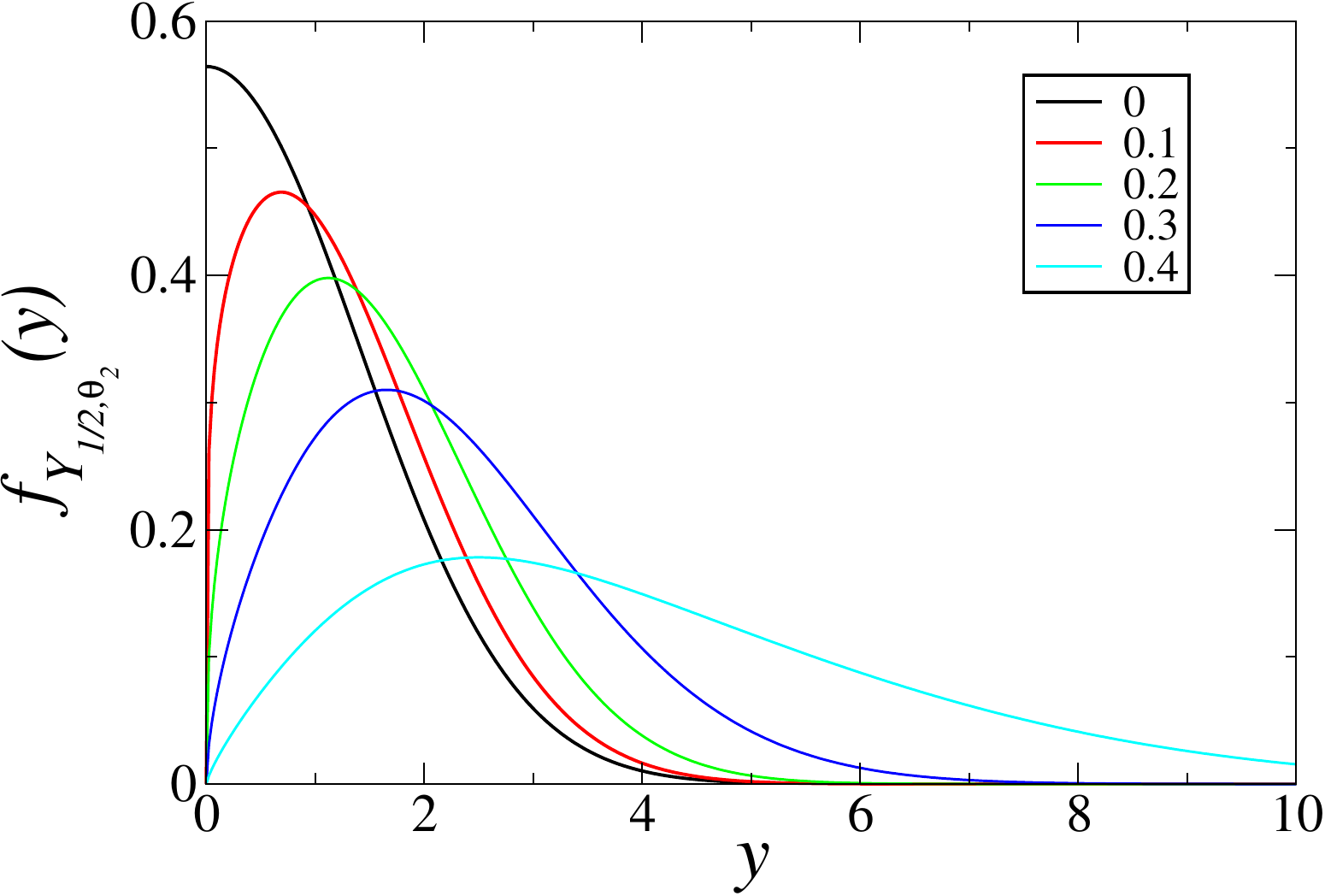}
\caption{\small
Probability density $f_{Y_{1/2,\th_2}}(y)$,
obtained by means of the integral representation~(\ref{eq:fint}),
using~(\ref{eq:12}) and~(\ref{eq:Rzeta+}),
for $\th_1=1/2$ and several values of $\th_2$ (see legend).}
\label{fig:half}
\end{center}
\end{figure}

The formulas~(\ref{eq:12}), (\ref{eq:hypers}) and~(\ref{eq:Rzeta+}) turn the
representation~(\ref{eq:fint})
into an efficient tool to evaluate the density $f_{Y_{1/2,\th_2}}(y)$ numerically.
Figure~\ref{fig:half} shows the distribution thus obtained,
for several values of $\th_2$ (see legend).
The general trend is that, as~$\th_2$ is increased from 0 to $\th_1$,
the density $f_{\Y}(y)$ broadens while its maximum shifts to the right.

To close, we mention that the representation~(\ref{eq:fint}) simplifies
whenever $\th_1=p/q$ is a rational number.
First, the density $f_{\X}(x)$ is related to the stable L\'evy law of index $\th$
(see~(\ref{eq:xleq})),
which is known to admit explicit expressions in terms of special functions
whenever $\th$ is rational~\cite{penson1,penson2}.
Furthermore, the series $\tilde I(\z)$ and $\tilde K(\z)$ can be expressed
as linear combinations of $q$ hypergeometric series,
by generalising the above construction.
The resulting expressions however soon become pretty cumbersome.

\section{Two special cases}
\label{sec:special}

\subsection{Poissonian internal process}
\label{sec:specialpoi}

When $\rho(\tau)$ is exponential, of the form
\beq\label{eq:rhoexp}
\rho(\tau)=\lam\e^{-\lam\tau},\qquad\hat\rho(s)=\frac{\lam}{\lam+s},
\eeq
the internal renewal process is a Poisson process,
implying, as shown below, that the statistics of $\nt$ is Poisson, too, regardless of the distribution $f(T)$ characterising the external process.
In the phase diagram of figure~\ref{fig:phase},
this case lies on the right boundaries of regions~A and~B, where $\th_1=\infty$.

For the internal process, the expression~(\ref{eq:Zzs}) simplifies to
\beq\label{eq:Poisson}
\hat Z(z,s)=\frac{1}{s+(1-z)\lam},
\eeq
hence
\beq\label{eq:genfish}
Z(z,t)=\e^{(z-1)\lam t},
\eeq
implying that the number $N_t$ of renewals up to time $t$ has
a Poisson distribution with parameter $\lam t$,
\beq\label{eq:fish}
\prob(N_t=n)=\e^{-\lam t}\frac{(\lam t)^n}{n!}.
\eeq

For the entire process, we have
\bea
\hat\phi(z,s)&=&\int_0^\infty\dd T\,f(T)\e^{-sT}\e^{(z-1)\lam T}
=\hat f(s+(1-z)\lam),
\nonumber\\
\hat\psi(z,s)&=&\int_0^\infty\dd T\,\Phi(T)\e^{-sT}\e^{(z-1)\lam T}
=\hat\Phi(s+(1-z)\lam)=\frac{1-\hat f\big(s+\lam(1-z)\big)}{s+\lam(1-z)},
\eea
and~(\ref{eq:key}) simplifies to
\be
\hat\Z(z,s)=\frac{1}{s+(1-z)\lam},
\ee
which is identical to~(\ref{eq:Poisson}), implying that $\nt$ has
a Poisson distribution~(\ref{eq:fish}) with parameter~$\lam t$,
regardless of the external density $f(T)$.
This result has the following interpretation.
As a consequence of~(\ref{eq:genfish}),
the numbers of points in the successive intervals $\T_1,\T_2,\dots,B_t$
have independent Poisson distributions,
with respective parameters $\lam\T_1,\lam\T_2,\dots,\lam B_t$,
implying that their sum~$\nt$ has a Poisson distribution whose parameter is the
sum
of all parameters, i.e., $\lam t$.

We have in particular, using the fact that $\lambda=1/\mean{\ttau}$,
\be
\mean{\nt}=\frac{t}{\mean{\ttau}}.
\ee
In region~A, this result is in agreement with (\ref{eq:NregionA}),
since $\mean{N_{\T}}=\mean{\T}/\mean{\ttau}$,
while for region~B, this result is precisely (\ref{eq:region3}).

\subsection{Poissonian resetting and dressed renewal process}
\label{sec:PoissonReset}

Poissonian resetting, which is the simplest---and the most studied---case of
stochastic resetting,
corresponds to the circumstance where $f(T)$ is exponential, of the form
\beq\label{eq:fexp}
f(T)=r\e^{-rT},
\eeq
hence $\Phi(T)=\e^{-rT}$.
In the phase diagram of figure~\ref{fig:phase},
this case lies on the upper boundary of region~A, where $\th_2=\infty$.

The general results of section~\ref{sec:nested} also simplify in this situation.
We have indeed
\be
\hat\phi(z,s)=r\hat Z(z,r+s),
\qquad\hat\psi(z,s)=\hat Z(z,r+s),
\ee
implying that~(\ref{eq:key}) takes on the familiar form
\beq\label{eq:renew}
\hat\Z(z,s)=\frac{\hat Z(z,r+s)}{1-r\hat Z(z,r+s)},
\eeq
which could alternatively have been obtained from a renewal equation%
\footnote{See~\cite{EMS} and references therein for similar relationships with other observables.}
and where $\hat Z(z,s)$ is given by~(\ref{eq:Zzs}).
Hence
\beq\label{eq:zedres}
\hat\Z(z,s)=\frac{1-\hat\rho(r+s)}{s+(r-(r+s)z)\hat\rho(r+s)}.
\eeq
When ${r}\to0$
the replication disappears and we consistently retrieve (\ref{eq:Zzs}).
By differentiating~(\ref{eq:zedres}) with respect to $z$ at $z=1$, we obtain in
particular
\be
\lap{t}\mean{\nt}=\frac{(r+s)\hat\rho(r+s)}{s^2(1-\hat\rho(r+s))},
\ee
entailing that, in the long-time regime,
\beq\label{eq:rescalN}
\mean{\nt}\approx\frac{r\hat\rho(r)}{1-\hat\rho(r)}\,t
=\frac{\hat\rho(r)}{1-\hat\rho(r)}\frac{t}{\mean{\T}},
\eeq
which is of the form~(\ref{eq:NregionA}) or (\ref{eq:Nnut}).

Remarkably enough, the expression~(\ref{eq:zedres}) is of the
form~(\ref{eq:Zzs}),
where $\hat\rho(s)$ is replaced by the expression
\beq\label{eq:rhors}
\hat\rho^{(\tt r)}(s)=\frac{(r+s)\hat\rho(r+s)}{s+r\hat\rho(r+s)}.
\eeq
We conclude that, in the present circumstance of Poissonian resetting,
the events of the internal process (the crosses in figure~\ref{fig:nested})
are exactly described by a single renewal process,
defined by the dressed density $\rho^{(\tt r)}(\tau)$,
depending on the resetting rate $r$ and on the distribution $\rho(\tau)$.
This is the  common probability density of the dressed interarrival times $\ttaur_1,\ttaur_2,\dots$, such that
\be
\hat\rho^{(\tt r)}(s)=\mean{\e^{-s\ttaur}}.
\ee
As $r\to0$, the dressed density $\rho^{(\tt r)}(\tau)$ reduces to the bare one $\rho(\tau)$, as expected.

Notice that, if one substitutes~(\ref{eq:Rs}) in~(\ref{eq:rhors}), the Laplace
transform of the dressed survival probability
$\hat R^{(\tt r)}(s)=(1-\rho^{(\tt r)}(s))/s$ takes on the familiar form
\be
\hat R^{(\tt r)}(s)=\frac{\hat R(r+s)}{1-r\hat R(r+s)},
\ee
 relating the expressions of the survival probability in the absence or in the presence of resetting (see, e.g.,~\cite{EMS}).
Considerations on the survival probability and the first-passage time in the
general case of an arbitrary resetting density $f(T)$ will be presented in
section~\ref{sec:first}.

In the particular case where $\rho(\tau)$ is the exponential
distribution~(\ref{eq:rhoexp}),
we recover $\rho^{(\tt r)}(\tau)=\rho(\tau)$,
irrespective of the rate $r$, in agreement with the results of
section~\ref{sec:specialpoi}.
The simplest non-trivial example is the case where $\rho(\tau)$
is the convolution of two exponentials, namely
\beq\label{eq:linexp}
\rho(\tau)=\lam^2\tau\e^{-\lam\tau},
\eeq
leading to
\be
\rho^{(\tt r)}(\tau)=\frac{\lam^2}{\omega}\,\e^{-(\lam+r/2)\tau}\sinh\omega\tau,\qquad
\omega=\frac{\sqrt{r(4\lam+r)}}{2}.
\ee

The dressed density $\rho^{(\tt r)}(\tau)$ has generically an exponential decay,
\be
\rho^{(\tt r)}(\tau)\sim\e^{-\mu\tau},
\ee
where the decay rate $\mu$
is the opposite of the nearest zero of the denominator
of~(\ref{eq:rhors})\footnote{In the example~(\ref{eq:linexp}),
$\mu=\lam+r/2-\omega$ decreases continuously from $\lam$ to 0 as $r$ is increased.},
obeying $\mu=r\hat\rho(r-\mu)$.
All moments of $\rho^{(\tt r)}(\tau)$ are therefore finite.
We have in particular (see~(\ref{eq:rescalN}))
\be
\mean{\ttaur}=\frac{1-\hat\rho(r)}{r\hat\rho(r)}.
\ee
This expression is expected to be related to $\mean{\ttau}$ at weak resetting.
More precisely,
when $\mean{\ttau}$ is finite, that is to say if $\th_1>1$,
then $\mean{\ttaur}$ does indeed converge to $\mean{\ttau}$ as $r\to0$.
Conversely, if $\mean{\ttau}$ is infinite, meaning $\th_1<1$,
$\mean{\ttaur}$ diverges as $r\to0$,
according to
\beq\label{eq:meantauP}
\mean{\ttaur}\approx\frac{\Gamma(1-\th_1)\tau_0^{\th_1}}{r^{1-\th_1}}.
\eeq
This expression is to be compared to the
marginal mean of a single time interval, given by~(\ref{eq:meantauR}), with $\th=\th_1<1$.
The two expressions~(\ref{eq:meantauR}) and~(\ref{eq:meantauP}) are similar,
with $1/r$ playing the role of the observation time.
In particular, both prefactors diverge in the $\th_1\to1$ limit.

\section{First-passage time under restart}
\label{sec:first}

In the case of Poissonian resetting (see section~\ref{sec:PoissonReset}),
the intervals of time $\ttaur_1,\ttaur_2,\dots$ between successive internal events (shown as crosses in figure~\ref{fig:nested}) are iid and drawn from the dressed density $\rho^{(\tt r)}(\tau)$.
This means that $\ttaur_2,\ttaur_3,\dots$ are probabilistic copies of the first
interval, $\ttaur_1$, which itself is the time of the first occurrence of a
renewal, or first-passage time for short, \textit{in the presence of resetting}.

In the general case where the external process has an arbitrary density $f(T)$,
there is no such renewal description for the sequence of crosses in terms of a dressed density.
Nevertheless, the time of the first occurrence of a renewal for the process with resetting,
or first-passage time, remains well defined, even if the
subsequent time intervals between crosses are no longer probabilistic
copies of this first interval.
Hereafter, we shall keep the same notations $\ttaur_1$ and $\rho^{(\tt r)}(\tau)$ to
refer to this first-passage time and its probability density.
The latter density can be derived in two complementary ways, as we now show.

We start from the observation that
\be
\mathcal{Z}(0,t)=\prob(\nt=0)=\prob(\ttaur>t)=\int_t^\infty\dd\tau\,\rho^{(\tt r)}(\tau),
\ee
hence, in Laplace space,
\be
\hat{\mathcal{Z}}(0,s)=\lap{t}\mathcal{Z}(0,t)
=\lap{t}\prob(\nt=0)=\frac{1-\hat\rho^{(\tt r)}(s)}{s}.
\ee
We now refer to~(\ref{eq:phidef}), (\ref{eq:psidef}) and~(\ref{eq:key}) to
obtain
\be
\hat{\mathcal{Z}}(0,s)
=\frac{\hat\psi(0,s)}{1-\hat\phi(0,s)},
\ee
with
\be
\hat\phi(0,s)
=\int_0^\infty\dd T\,\e^{-sT}f(T)Z(0,T)
=\int_0^\infty\dd T\,\e^{-sT}f(T)R(T)
\ee
and
\be
\hat\psi(0,s)
=\int_0^\infty\dd T\,\e^{-sT}\Phi(T)R(T).
\ee
Put together, the above equations lead to the desired result
\beq\label{eq:firstpas}
\hat\rho^{(\tt r)}(s)=\mean{\e^{-s\ttaur_1}}
=\frac{\int_0^\infty\dd T\,\e^{-sT}\rho(T)\Phi(T)}{1-\int_0^\infty\dd T\,
\e^{-sT}f(T)R(T)}.
\eeq
The identity
\be
\int_0^\infty\dd T\bigl(\rho(T)\Phi(T)+f(T)R(T)\bigr)=1,
\ee
which simply states that $\prob(\ttau<\T)+\prob(\ttau>\T)=1$,
allows one to check that $\hat\rho^{(\tt r)}(0)=1$,
i.e., that the density $\rho^{(\tt r)}(\tau)$
is normalised.
In the particular case where $f(T)$ is exponential
(see~(\ref{eq:fexp})),~(\ref{eq:firstpas}) becomes~(\ref{eq:rhors}), as it should~be.

Differentiating~(\ref{eq:firstpas}) with respect to $s$ at $s=0$ yields
the following expression for the mean first-passage time,
\be
\mean{\ttaur_1}=\frac{\int_0^\infty\dd T R(T)\Phi(T)}{\int_0^\infty\dd T \rho(T)\Phi(T)},
\ee
which is well defined when the tail exponents obey the inequality $\th_1+\th_2>1$.
Interestingly enough,
this inequality does not show up in the construction
of the phase diagram of figure~\ref{fig:phase}.
This discrepancy lies in the fact that the nature of $\ttaur_1$ is that of a `boundary' observable,
whereas the phase diagram concerns the `bulk' observable $\nt$.
The same observation applies for the statistics of the `boundary' observable $N_{B_t}$ (see section~\ref{sec:NBt}).

In order to better understand the meaning of~(\ref{eq:firstpas}),
we present an alternative derivation, inspired by~\cite{reuveniPRL,reuveni,bonomo}.
The first-passage time $\ttaur_1$ obeys the recursion
\beq\label{eq:branches}
\ttaur_1=\left\{
\begin{array}{cc}
\ttau\hfill &\hbox{if $\ttau<\T$},\cr
\T+(\ttaur_1)^\prime\quad &\hbox{if $\ttau>\T$},
\end{array}
\right.
\eeq
where $(\ttaur_1)^\prime$ is an independent copy of $\ttaur_1$.
Let
\be
p=\prob(\ttau<\T)=\int_0^\infty\dd T\,\rho(T)\Phi(T).
\ee
Iterating~(\ref{eq:branches}), we obtain
\be
\ttaur_1=\tilde\T_1+\cdots+\tilde\T_k+\tilde\ttau_{k+1},
\ee
with probability $p(1-p)^k$,
where the tilde indicates that these random variables are conditioned by the
inequalities on the right-hand side of~(\ref{eq:branches}).
We thus infer that
\be
\hat\rho^{(\tt r)}(s)=\sum_{k\ge0}p(1-p)^k
\hat f_{\tilde\T}(s)^k\,f_{\tilde\ttau}(s)
=\frac{p f_{\tilde\ttau}(s)}{1-(1-p)\hat f_{\tilde\T}(s)},
\ee
where
\be
f_{\tilde\ttau}(\tau)=\frac{\rho(\tau)\Phi(\tau)}{p},
\qquad f_{\tilde\T}(T)=\frac{f(T)R(T)}{1-p}.
\ee
which, again, lead to~(\ref{eq:firstpas}).
We refer to~\cite{reuveniPRL,reuveni,bonomo} for further considerations on the topic of
first passage under restart.

Yet another derivation of~(\ref{eq:firstpas}), based on a renewal equation for the survival probability 
$\prob(\ttaur_1>t)$, can be found in~\cite{EMS} (cf references therein).

\section{Number of internal renewals $N_{B_t}$ in the last interval}
\label{sec:NBt}

As mentioned earlier, nested renewal processes have been initially introduced
in the context of reliability
problems~\cite{ansell,bendell,bendell-b,degbotse}.
Studying the statistics of $N_{B_t}$ was one of the primary aims
of~\cite{ansell}.
This quantity represents the number of internal renewals in the interval $B_t$,
which is the backward recurrence time for the external renewal process (see
figure~\ref{fig:nested}).
In~\cite{ansell}, the study was restricted to the case of thin-tailed
distributions for both the internal and external renewal processes, where the
statistics of $N_{B_t}$ becomes stationary at long times.
Here we shall be interested in exploring the phase diagram in the whole $\th_1$--$\th_2$-plane.

\subsection{On the statistics of $N_{B_t}$}

The statistics of $N_{B_t}$ can be determined by utilising~(\ref{eq:B}),
(\ref{eq:probNB}), (\ref{eq:probNBk}) and (\ref{eq:NB}).

\begin{figure}
\begin{center}
\includegraphics[angle=0,width=.75\linewidth,clip=true]{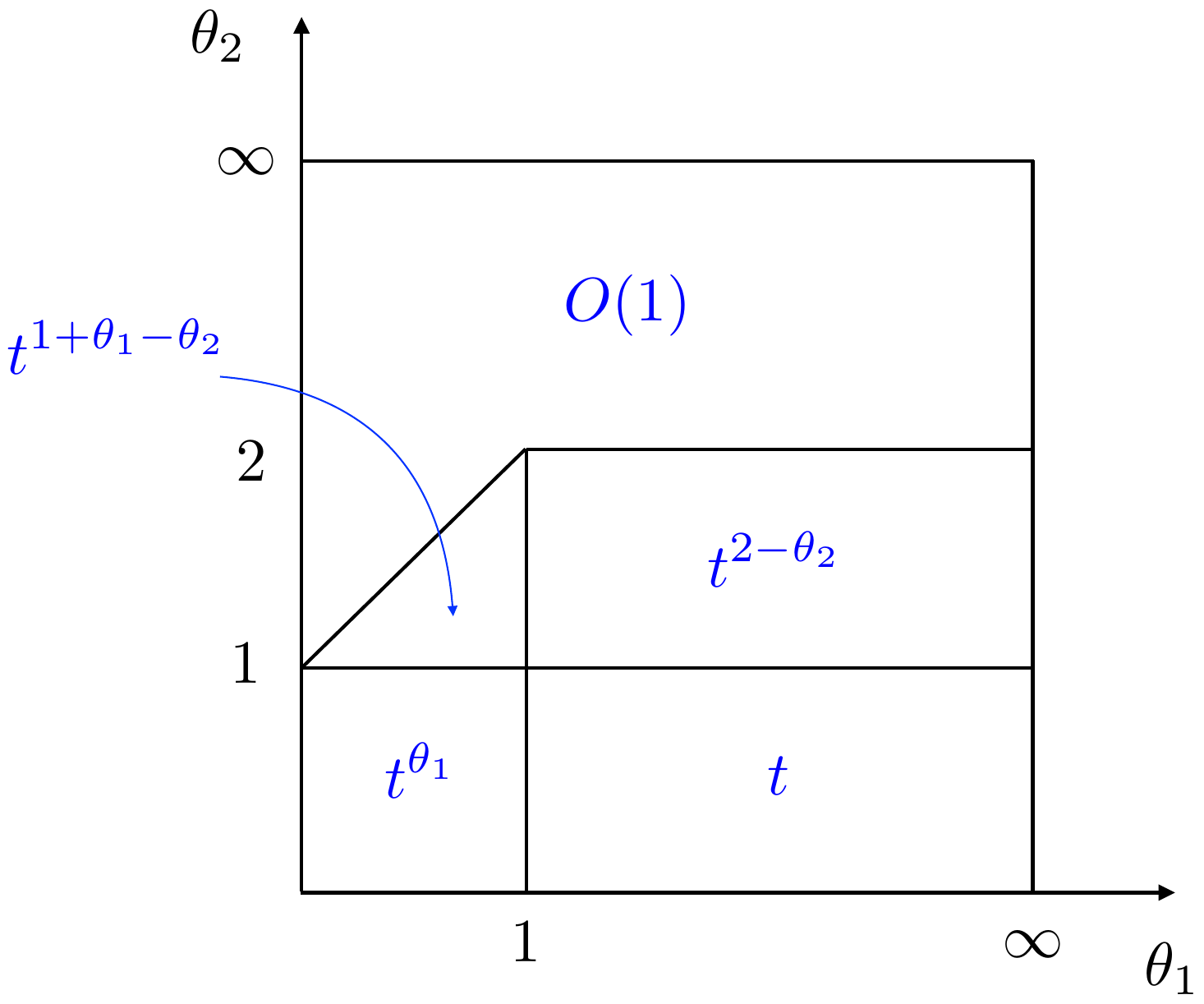}
\caption{\small
Asymptotic behaviour of the mean number of internal renewals $\mean{N_{B_t}}$ in
the interval $B_t$, the backward recurrence time of the external process.}
\label{fig:NB}
\end{center}
\end{figure}

Let us focus on the asymptotic behaviour of its mean $\mean{N_{B_t}}$.
Figure~\ref{fig:NB} gives a summary of the results obtained
using techniques similar to those outlined in section~\ref{sec:diagram}, as elaborated below.

Beforehand, to gain an intuitive understanding of this phase diagram, 
one can compare the mean values of the two time intervals: $B_t$ for the resetting process, and $\ttau_t$ for the internal renewal process.
For the former, we have~\cite{glrenew}
\be
\mean{B_t}\sim
\begin{cases}
O(1),& \th_2>2,\\
t^{2-\th_2},& 1<\th_2<2,\\
t,& \th_2<1,
\end{cases}
\ee
while (see~(\ref{eq:meantauR}))
\be
\mean{\ttau_t}\sim
\begin{cases}
O(1),& \th_1>1,\\
t^{1-\th_1},& \th_1<1,
\end{cases}
\ee
for the latter.
For instance, when both exponents are greater than unity, 
$\mean{N_{B_t}}\sim\mean{B_t}/\mean{\ttau_t}\sim O(1)$,
when both exponents are less than unity, $\mean{N_{B_t}}\sim
\mean{B_t}/\mean{\ttau_t}\sim t^{\th_1}$, and so on.

A more detailed approach is as follows.
The number of renewals $N_{B_t}$ in the random interval $B_t$ can be expressed, asymptotically, as (cf section~\ref{sec:Ntas})
\beq\label{eq:NBtas}
N_{B_t}\approx
\begin{cases}
\frad{B_t}{\taum},&\th_1>1,
\vspace{2pt}
\\
\frad{X_{\th_1}}{\Gamma(1-\th_1)}\left(\frad{B_t}{\tau_0}\right)^{\th_1},&\th_1<1
\end{cases}
\eeq
where, asymptotically~\cite{glrenew},
\beq\label{eq:fBtas}
B_t\approx
\begin{cases}
B_\eq,&\th_2>1,
\\
t\beta,&\th_2<1.
\end{cases}
\eeq
Hence, if $\th_2>1$, the probability density $f_{B_t}(t,B)$ of the backward recurrence time $B_t$ is asymptotically given by
\be
f_{B,{\eq}}(B)=\frac{\Phi(B)}{\mean{\T}}
\approx\frac{1}{\mean{\T}}
\left(\frac{T_0}{B}\right)^{\th_2},
\ee
while, if $\th_2<1$, it is given in terms of the density of the rescaled random variable $\beta$
\be
f_\beta(x)=
\frac{\sin\pi\th}{\pi}x^{-\th}(1-x)^{\th-1}=\beta_{1-\th,\th}(x)
\qquad (0<x<1),
\ee
where
\be
\beta_{a,b}(x)=\frac{\Gamma (a+b)}{\Gamma (a)\Gamma (b)}x^{a-1}(1-x)^{b-1} 
\ee
is the beta distribution on $[0,1]$~\cite{glrenew}. 

The asymptotic expressions of $\mean{N_{B_t}}$ follow readily from~(\ref{eq:NBtas}) and~(\ref{eq:fBtas}), using~(\ref{eq:NB}) or, equivalently, in direct space,
\be
\mean{N_{B_t}}=\int_0^t\dd B\,f_{B_t}(t,B)\mean{N_B}.
\ee
This expression is the parallel of expression~(\ref{eq:NCox}).

Thus, if $\th_1>1$, 
\be
\mean{N_{B_t}}
\approx
\begin{cases}
\frad{\mean{\T^2}}{2\taum\mean{\T}},&\th_2>2,
\vspace{2pt}
\\
\frad{T_{0}^{\th_2}}{(2-\th_2)\taum\mean{\T}}t^{2-\th_2},&1<\th_2<2,
\vspace{2pt}
\\
\frad{1-\th_2}{\taum}t,&\th_2<1.
\end{cases}
\ee
If $\th_1<1$, we have
\be
\mean{N_{B_t}}
\approx
\begin{cases}
\frad{I_2(0)}{\mean{\T}},&1+\th_1<\th_2,
\vspace{2pt}
\\
\frad{\sin\pi\th_1\,T_0^{\th_2}}{\pi\th_1\,\tau_0^{\th_1}\,\mean{\T}}\frad{t^{1+\th_1-\th_2}}{1+\th_1-\th_2},&1<\th_2<1+\th_1,
\vspace{3pt}
\\
\frad{\sin\pi\th_1}{\pi\th_1}\mean{\beta^{\th_1}}\left(\frad{t}{\tau_0}\right)^{\th_1},&\th_2<1.
\end{cases}
\ee
The constant numerator appearing in the first line is the finite limit of $I_2(s)$ when $s\to0$,
\be
I_2(0)=\int_0^\infty\dd T\,\Phi(T)\mean{N_T}.
\ee

Finally, notice that $N_{B_t}$ contributes to the total sum $\nt$, given in~(\ref{eq:Ntot}), in regions B and D only.
However, in both regions, its behaviour differs from that of $\nt$.
In region B, $\nt$ has negligible fluctuations around its mean, while $N_{B_t}\sim t\beta$ keeps fluctuating, which means that the fluctuations of the sum of the $M_t$ first terms in~(\ref{eq:Ntot}) compensate those of $N_{B_t}$.
In region D, $\nt\sim \Y\,t^{\th_1}$, while $N_{B_t}\sim X_{\th_1}\beta^{\th_1}t^{\th_1}$, meaning that all the complexity of the behaviour of $\nt$ lies in the sum of the $M_t$ first terms in~(\ref{eq:Ntot}).

\subsection{Continuous time random walk subject to resetting}

As mentioned earlier (see section~\ref{sec:over}), a continuous time random walk subject to resetting involves two nested renewal processes.
The process considered in~\cite{ansell}, and recalled in section~\ref{sec:intro}, is equivalent to a continuous time random walk, where the shocks, causing damages of magnitude $\eta_1,\eta_2,\dots$, with common density $f_\eta$, correspond to the jumps.
The cumulative damage of the component in use is to be identified with the position of the walker at time $t$, that is,
\be
X_t=\eta_1+\eta_2+\cdots+\eta_{N_{B_t}},
\ee
with probability density
\be
f_{X_t}(x)=\sum_{n\ge0}(f_\eta\star)^n(x)\,\prob(N_{B_t}=n).
\ee
Assuming, for instance, that the distribution of the steps $\eta_1,\eta_2,\dots$ is symmetric, one easily finds that the mean squared displacement of the walker reads
\be
\mean{X_t^2}=\mean{N_{B_t}}\mean{\eta^2}.
\ee
The computation of the mean squared displacement for a continuous time random walk under power-law resetting has previously been addressed in~\cite{bodrova}, resulting in a phase diagram for the asymptotic time dependence of this quantity in the $\th_1$--$\th_2$-plane.
This phase diagram corresponds precisely to the one depicted in figure~\ref{fig:NB}.

\bmhead{Acknowledgments}
It is a pleasure to thank Pierre Vanhove for an interesting discussion.

\vsk
{\bf Data availability statement.}
The authors have no data to share.

\vskip6pt\noindent
{\bf Conflict of interest.}
The authors declare no conflicts of interest.

\bibliography{paperNested}

\end{document}